\documentclass[twocolumn, 10pt, aps, showpacs, reprint, amsmath, amssymb, superscriptaddress, nofootinbib, floatfix]{revtex4-2}

\usepackage[]{graphicx}
\usepackage{xspace}

\newcommand{\Lim}{Lima\c{c}on\xspace}
\newcommand{\neff}{n_\mathrm{eff}}
\newcommand{\Qmax}{Q_\mathrm{max}}
\renewcommand{\Re}{\mathrm{Re}}
\renewcommand{\Im}{\mathrm{Im}}

\begin{document}
	
	\title{Impact of cavity geometry on microlaser dynamics}
	\author{Kyungduk Kim$^*$}
	\affiliation{Department of Applied Physics, Yale University, New Haven, Connecticut 06520, USA}
	\author{Stefan Bittner$^*$}
	\affiliation{Universit\'e de Lorraine, CentraleSup\'elec, LMOPS, 2 rue Edouard Belin, Metz 57070, France}
	\affiliation{Chair in Photonics, CentraleSup\'elec, LMOPS, 2 rue Edouard Belin, Metz 57070, France}
	\author{Yuhao Jin}
	\author{Yongquan Zeng}
	\author{Qi Jie Wang}
	\affiliation{Center for OptoElectronics and Biophotonics, School of Electrical and Electronic Engineering, School of Physical and Mathematical Science, and Photonics Institute, Nanyang Technological University, 639798, Singapore}
	\author{Hui Cao$^\dagger$}
	\affiliation{Department of Applied Physics, Yale University, New Haven, Connecticut 06520, USA}
	
	\begin{abstract}
		We experimentally investigate spatio-temporal lasing dynamics in semiconductor microcavities with various geometries, featuring integrable or chaotic ray dynamics. The classical ray dynamics directly impacts the lasing dynamics, which is primarily determined by the local directionality of long-lived ray trajectories. The directionality of optical propagation dictates the characteristic length scales of intensity variations, which play a pivotal role in nonlinear light-matter interactions. While wavelength-scale intensity variations tend to stabilize lasing dynamics, modulation on much longer scales causes spatial filamentation and irregular pulsation. Our results will pave the way to control the lasing dynamics by engineering the cavity geometry and ray dynamical properties.
	\end{abstract}

	\maketitle
	
	Controlling nonlinear dynamics of complex systems is crucial in, e.g., nonlinear optics, hydrodynamics, and laser physics~\cite{ott1990controlling,roy1992dynamical}. It remains, however, a challenge to control semiconductor laser dynamics due to extremely fast inherent time scales \cite{OhtsuboBook2013}. Instead of relying on external feedback~\cite{mork1990route,fischer1994high,fischer1996fast,martin1996mode,mandre2005spatiotemporal} or optical injection~\cite{wieczorek1999unifying,hwang2000dynamical,krauskopf2000different,takimoto2009control}, we propose a more direct and compact approach based on modifying the intrinsic light-matter interaction inside the laser by tailoring the cavity geometry~\cite{Bittner2018a, Kim2021, Kim2022}.  
	
	Broad-area Fabry-Perot cavities, commonly used for semiconductor lasers, often result in lasing instabilities~\cite{Fischer1996, Hess1996, Marciante1997, Marciante1998, Klaedtke2006, Scholz2008, Arahata2015}. An optical lensing effect in high-intensity regions, caused by spatial hole-burning and carrier-induced refractive index changes, leads to self-focusing of light and the formation of filaments, which are unstable and induce irregular pulsations. 	
	We recently showed that modifying the resonator shape can suppress spatio-temporal instabilities, e.g., in a D-shaped microcavity that features fully chaotic ray dynamics~\cite{Bittner2018a}, or in a stable cavity with optimized mirror curvature that exhibits integrable ray dynamics~\cite{Kim2021, Kim2022}. 
	The stability of lasing dynamics depends critically on the characteristic length scale of optical intensity variations in the cavity. If high-intensity regions, formed by constructive interference of propagating waves, are too small to induce a lensing effect, filamentation and lasing instability will be prevented. However, it is not clear how the cavity geometry affects the characteristic size of high-intensity regions, and whether it is possible to predict the lasing dynamics based on classical ray dynamics. 

	To address these questions, we experimentally study how the classical ray dynamics is related to the spatio-temporal dynamics of semiconductor microcavity lasers. The ray dynamics of two-dimensional (2D) optical microcavities is entirely determined by the cavity shape and the boundary conditions. Hence, microlasers based on dielectric resonators correspond to billiards with (partial) ray escape according to Fresnel's laws \cite{Altmann2013, Cao2015}. Previously the ray-wave correspondence was studied to reveal the relation between long-lived trajectories and resonances with high quality (Q) factors~\cite{chang1996optical, stone2001wave, Tureci2005, harayama2011two, Cao2015}. Most studies have concentrated on static properties like emission spectra and far-field distributions~\cite{levi1993directional, nockel1996directional, chern2003unidirectional, Schwefel2003, Schwefel2004, Fukushima2005}, except a few numerical \cite{Sunada2005, Harayama2007a} and experimental studies \cite{Choi2008a, Shinohara2014, Bittner2018a, Ma2022a, kim2023spatiotemporal} of the dynamic properties of deformed microcavity lasers. 
	
	Here we extend the paradigm of ray-wave correspondence to engineering the nonlinear lasing dynamics by utilizing classical ray dynamics. We judiciously choose several cavity shapes not only with distinct types of ray dynamics: chaotic vs. integrable, but more importantly, with varying structural sizes and degrees of spatial localization of their lasing modes. We find the strength and occurrence of irregular pulsations of the laser emission are strongly correlated with the local structure size of the lasing modes, which are normally the most long-lived passive cavity resonances. This characteristic size is determined by the local directionality of optical propagation inside the cavity, which can be predicted by the classical ray dynamics. 
	
	Figure 1 shows five cavity geometries, D-cavity, stadium, \Lim, ellipse, and square. These five shapes include cavities with chaotic (D-cavity, stadium, \Lim) and integrable (ellipse, square) ray dynamics. Furthermore, D-cavity, stadium, and square have spatially extended modes featuring small structure size due to low local directionality of wave propagation,  while \Lim and ellipse have spatially localized whispering gallery modes (WGMs) exhibiting large structure size due to directional wave propagation.

	\begin{figure}[t]
		\includegraphics[width = 8.6 cm]{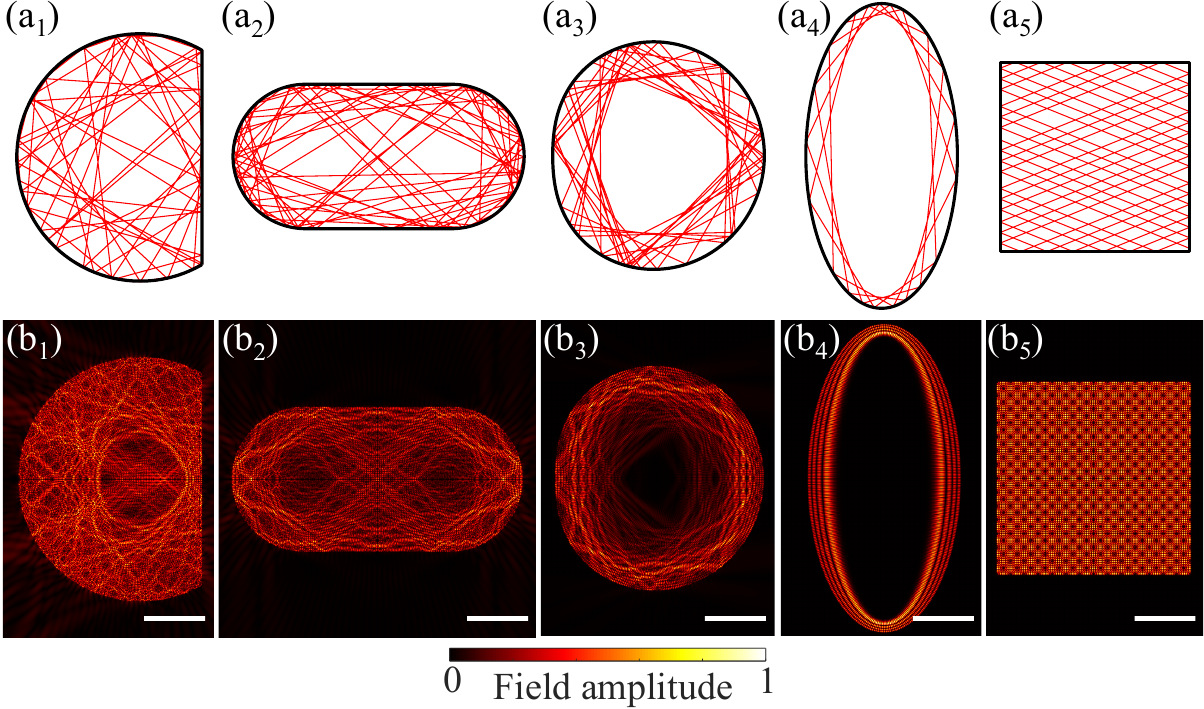}
		\caption{Five cavity geometries with different classical ray dynamics. (a) Typical long-lived ray trajectories in D-shaped (a$_1$), stadium (a$_2$), \Lim (a$_3$), ellipse (a$_4$), and square (a$_5$) dielectric cavities. (b) Numerically calculated mode profile with high Q-factor in a cavity with smooth boundary. The scale bars are $5~\mu$m long.}
		\label{fig1}
	\end{figure}
	
	D-cavity and stadium have fully chaotic ray dynamics~\cite{Bunimovich1979, Ree1999}, and long-lived rays explore the bulk of the cavities [Figs.~\ref{fig1}(a$_1$, a$_2$)]. Due to ray-wave correspondence~\cite{Berry1977, Shnirelman1974, Altmann2013, ketzmerick2022chaotic,you2022universal}, the high-Q modes are spatially extended and irregularly structured [Figs.~\ref{fig1}(b$_1$, b$_2$)]. The \Lim-shaped cavity features predominantly chaotic ray dynamics~\cite{Robnik1983,Wiersig2008,Dullin2001}, but has long-lived trajectories concentrated at the cavity boundary [Fig.~\ref{fig1}(a$_3$)], resulting in irregular WGMs [Fig.~\ref{fig1}(b$_3$)]. The ellipse features integrable ray dynamics \cite{Berry1981} with long-lived trajectories along the cavity boundary that are confined by total internal reflection [Fig.~\ref{fig1}(a$_4$)], resulting in regularly-structured WGMs [Fig.~\ref{fig1}(b$_4$)]. The square also features integrable ray dynamics, but its long-lived trajectories explore the whole cavity [Fig.~\ref{fig1}(a$_5$)], and hence its high-Q resonances~\cite{Bittner2013b, Bittner2014a, Yang2016} are spatially extended with regular, fine features [Fig.~\ref{fig1}(b$_5$)].
	
	We fabricate edge-emitting semiconductor microlasers with these five resonator shapes~\cite{SM}. The devices are fabricated on a commercial laser diode wafer with a GaAs/AlGaAs quantum well (Q-Photonics QEWLD-808) by photolithography and inductively coupled plasma etching, followed by deposition of a top metal contact for electric current injection. Multiple cavities for each of the five geometries, with identical cavity areas ($2.53\times10^4~\mu\mathrm{m}^2$), are fabricated on the same wafer.
	
	\begin{figure*}[t]
		\includegraphics[width = 15 cm]{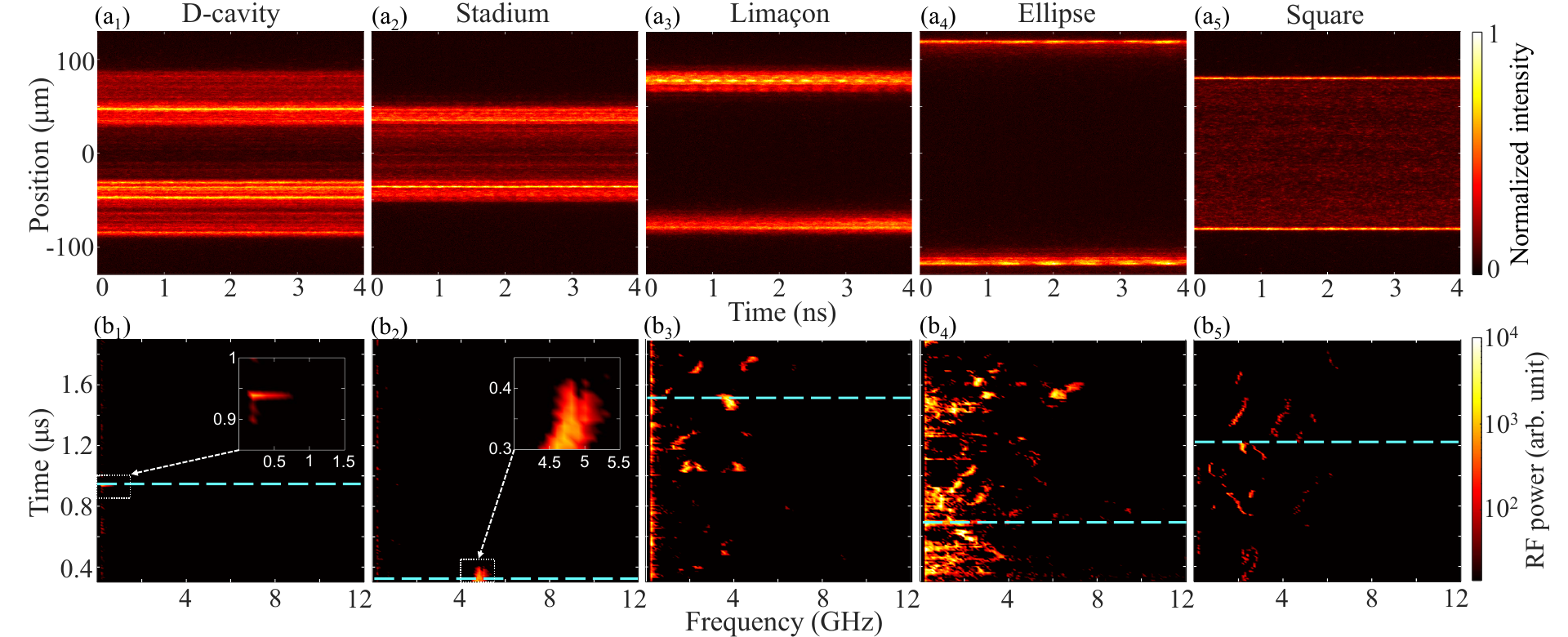}
		\caption{Experimentally measured spatio-temporal lasing dynamics. (a$_1$-a$_5$) Streak images of lasing emission from five different cavity geometries. The intensity is normalized by its maximum. The pump current is 500 mA for all cavities, well above their lasing thresholds. Intensity pulsations indicate unstable lasing dynamics. (b$_1$-b$_5$) Time-resolved RF-spectra $\hat{S}(f,t_d)$ of laser emission intensities obtained from the streak images. Insets in (b$_1$,b$_2$) are the magnifications of the white boxes. The cyan dashed lines indicate the times of the images in (a). RF peaks (bright spots) correspond to intensity pulsations.}
		\label{fig2}
	\end{figure*}
	
	We study their spatio-temporal lasing dynamics experimentally using a streak camera (Hamamatsu C5680/M5676) to record the time-resolved near-field emission intensity profiles. We measure 10 ns long time windows with a resolution of $\sim$30 ps. Lasing occurs in all cavities with electrical pumping at room temperature~\cite{SM}. Figure~\ref{fig2}(a) shows exemplary streak images. The spatial profiles of emission intensity agree with the output patterns of high-Q resonances~\cite{Bittner2020,kim2023spatiotemporal}. The irregular pulsations with typical periods of sub-nanoseconds are caused by unstable lasing dynamics. 
	
	We analyze the laser intensity fluctuations using the short-term radio-frequency (RF) spectra and their temporal evolution. We calculate the temporal Fourier transform of every 10-ns-long streak image and average its magnitude squared in space. Then a broadband continuous signal from the spatio-temporal beating of lasing modes and amplified spontaneous emission is subtracted from the RF spectra, which highlights discrete peaks from irregular pulsations due to lasing instabilities \cite{SM}. The subtracted RF-spectra in Figs.~\ref{fig2}(b$_1$-b$_5$) clearly show different degrees of stability for the five geometries.
	
\begin{figure}[b]
		\includegraphics[width = 8 cm]{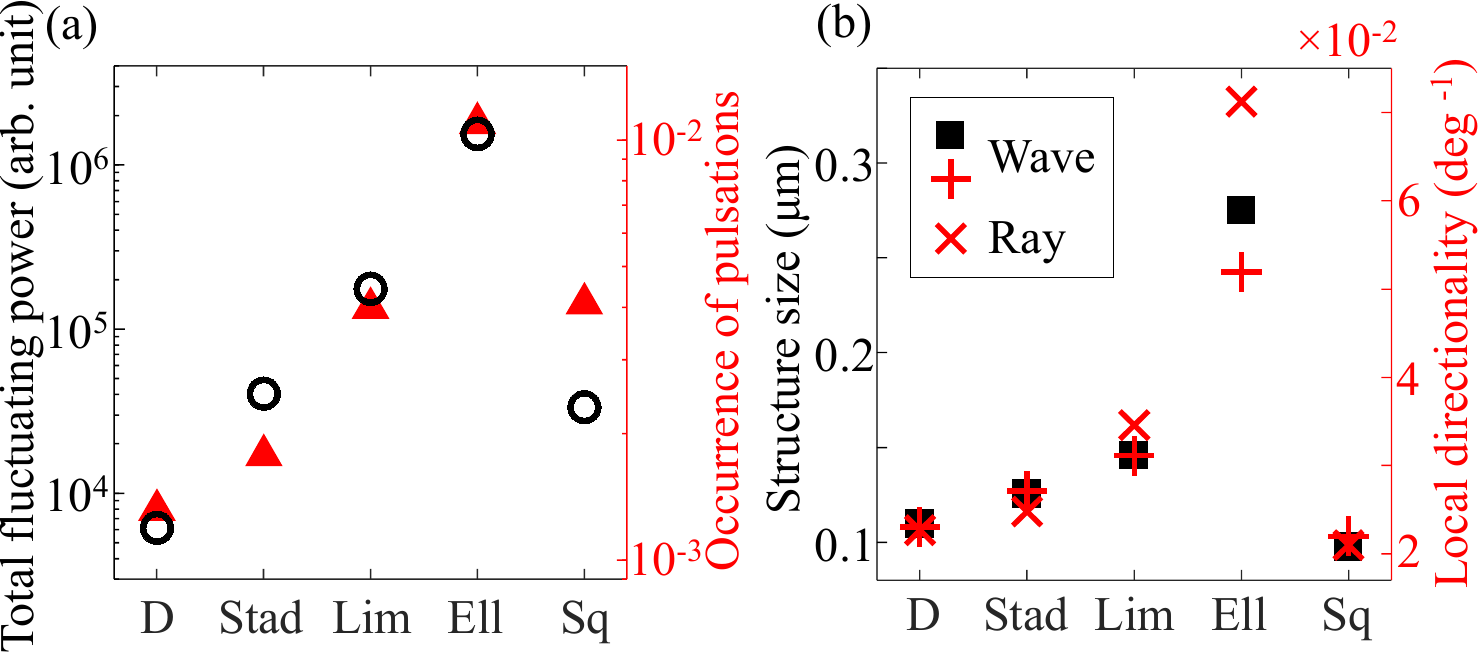}
		\caption{Relation between lasing instabilities and spatial structure of lasing modes. (a) Experimentally measured total RF-power $S_\mathrm{tot}$ of intensity pulsations (black circles, left axis) and the frequency of their occurrence (red triangles, right axis) for different cavity shapes. The results of five lasers per geometry are averaged on a logarithmic scale. (b) Numerically calculated local structure size $\langle s \rangle$ in five cavities with boundary roughness, averaged over the cavity area (black squares, left axis). It is determined by the local directionality of light propagation $\langle D_W \rangle$ and $\langle D_R \rangle$ (red symbols, right axis), obtained from wave ($+$) and ray simulations ($\times$).}
		\label{fig3}
	\end{figure}

	For a more quantitative statistical analysis, we characterize the RF-spectra with two measures in Fig.~\ref{fig3}(a). The first one (left axis) is the total RF power $S_{\mathrm{tot}}$, obtained by integrating the short-term RF spectra [Fig.~\ref{fig2}(b)] in both frequency and time. The second one (right axis) measures the frequency of occurrence of RF peaks by the participation ratio of the RF spectra \cite{SM}. To account for the cavity-to-cavity variations, we average these quantities over five different lasers per geometry. Both the overall fluctuation power and the occurrence of intensity pulsations vary by several orders of magnitude for different cavity geometries. The D-cavity microlasers have the weakest and rarest pulsations, followed by the stadia. \Lim cavities have stronger and more frequent pulsations than D-cavities and stadia. The ellipse lasers have the strongest and most frequent pulsations. The square cavities, in contrast, are much more stable than ellipses and close to stadia.
	
	The spatial structure of the lasing modes strongly influences the nonlinear interaction between the optical field and gain material, which in turn affects the strength and occurrence of irregular pulsations. To reveal the underlying mechanism, we numerically characterize the fine structure of the lasing modes. We calculate the passive resonances of 2D cavities with boundary roughness to account for fabrication defects \cite{SM}. The cavity dimensions are 10 times smaller than the actual ones to reduce computational load. Furthermore, we employ steady-state ab-initio lasing theory with single-pole approximation (SPA-SALT) to determine which cavity resonances will lase and to calculate their lasing intensities~\cite{SM, Ge2010, liew2015pump, Cerjan2016, Cerjan2019}.
	
	\begin{figure*}[ht]
		\includegraphics[width = 16 cm]{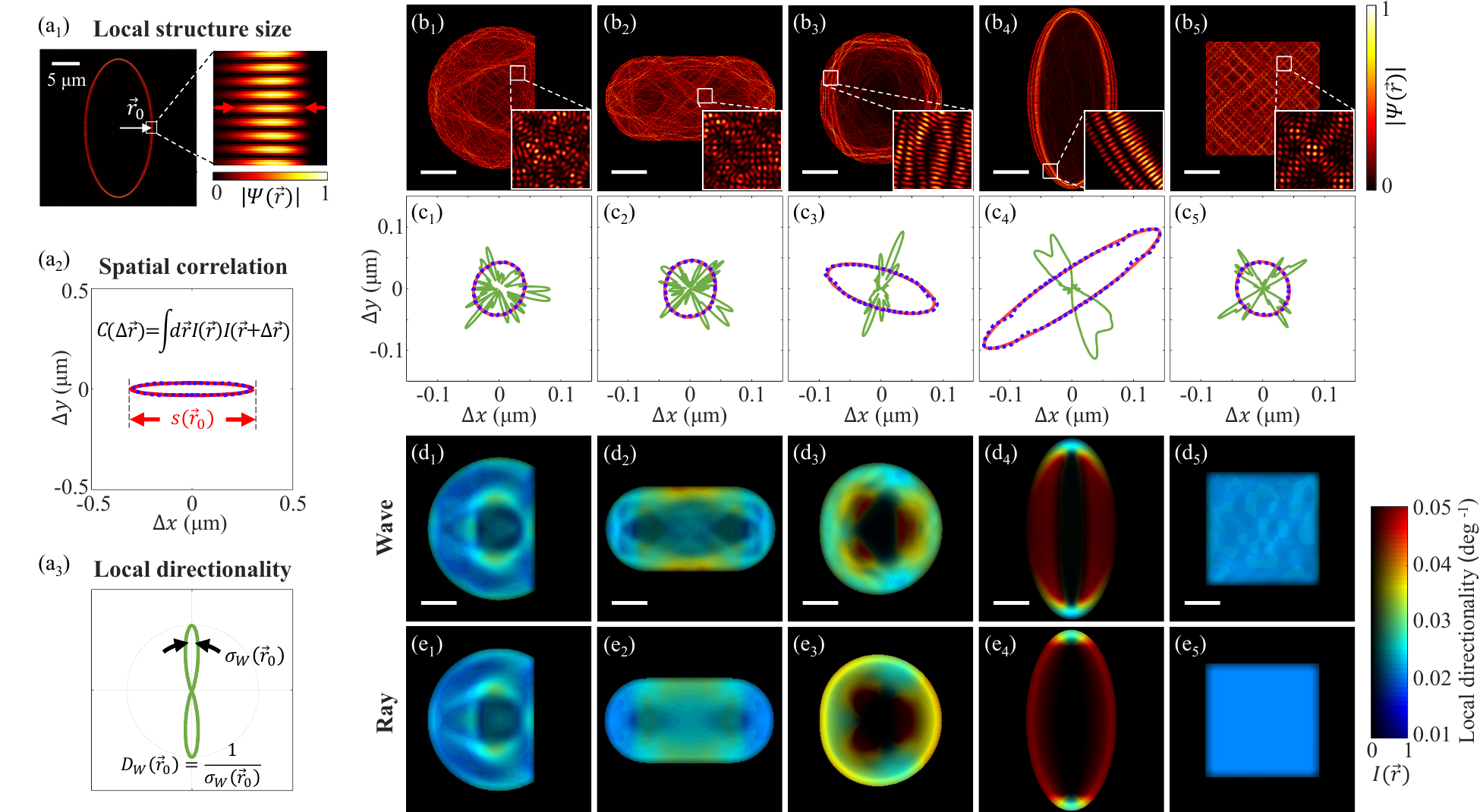}
		\caption{Numerically simulated local directionality of optical propagation. (a$_1$-a$_3$) Illustration of local structure size and local directionality. (a$_1$) An exemplary high-Q mode in an elliptic cavity features highly directional optical propagation, which yields elongated bright regions. The red arrows denote the structure size. (a$_2$) The half-maximum contour line of the spatial intensity correlation function $C(\Delta \vec{r})$ (blue dotted lines) for the white box centered at $\vec{r}_0$ in (a$_1$). It is fitted by an ellipse (solid red line) whose major axis is equal to the elongated feature size in (a$_1$) and defines the local structure size of 0.6 $\mu$m (red arrows). (a$_3$) Polar plot of the amplitude of local Fourier component $|w_\mu(\vec{r}_0, \theta)|$ (green), obtained by wavelet transform at the center $\vec{r}_0$ of the white box in (a$_1$). The standard deviation of its amplitude squared is 6.7$^\circ$ (black arrows), whose inverse yields the local directionality. 
		(b$_1$-b$_5$)~Typical high-Q modes of five cavity shapes with surface roughness. The scale bars are 5 $\mu$m long. (Insets)~Magnifications of the white boxes. 
		(c$_1$-c$_5$)~The analysis illustrated in (a) is performed with the modes in (b). 
		(d$_1$-d$_5$)~Maps of local directionality $D_W(\vec{r})$ of the simulated lasing modes. The color denotes the local directionality (red: high, blue: low), and the brightness denotes the local intensity of lasing modes. 
		(e$_1$-e$_5$)~Map of local directionality $D_R(\vec{r})$ obtained from ray tracing, showing excellent agreement with wave simulations in (d).}
		\label{fig4}
	\end{figure*}
	
	The mode profiles in Fig.~\ref{fig4}(b) show significant differences in their fine structure for the different geometries. The D-cavities and stadia [Figs.~\ref{fig4}(b$_1$,b$_2$)] show random and isotropic intensity variations on the scale of the in-medium wavelength. For the \Lim [Fig.~\ref{fig4}(b$_3$)], in contrast, the fine structure is anisotropic with elongated high-intensity grains. This anisotropy is even more pronounced for the ellipses [Fig.~\ref{fig4}(b$_4$)]. In contrast, the squares [Fig.~\ref{fig4}(b$_5$)] feature a more regular structure with a feature size similar to that of D-cavities or stadia. 
	
	To characterize the typical size of the fine structure, we compute the spatial intensity correlation functions of high-Q resonances in a local area \cite{SM}. The contour lines at half-maximum of the intensity correlation functions [blue dashed lines in Fig.~\ref{fig4}(c)] are fitted by an ellipse [red solid lines in Fig.~\ref{fig4}(c)]. While the length of the minor axis is consistent for all cavity shapes, the major axis of the ellipse varies significantly with the cavity geometry, and it is defined as the local structure size $s(\vec{r})$.
	
    We calculate the average structure size $\langle s \rangle$ by first averaging over all lasing modes weighted by their intensities, and then averaging over all spatial locations weighted by local intensity~\cite{SM}. Figure~\ref{fig3} shows that $\langle s \rangle$ [Fig.~\ref{fig3}(b), left axis] is strongly correlated with the lasing instabilities in the experiment [Fig.~\ref{fig3}(a)]. The irregular pulsations of broad-area semiconductor lasers originate from carrier-induced modulational instability. In a GaAs quantum well, high optical intensity depletes the local gain by spatial hole burning, which increases the refractive index locally. The resulting optical lensing effect and self-focusing lead to the formation of spatial filaments, which are inherently unstable and cause pulsations~\cite{Hess1996, Marciante1997, Marciante1998}. For cavities with $\langle s \rangle \sim \lambda$, intensity variation on a wavelength scale causes a refractive index change on the same scale, which is too small to focus light~\cite{Bittner2018a}, thus preventing filamentation and instability. Conversely, a large feature size $\langle s \rangle \gg \lambda$ is more likely to create a lensing effect, which leads to stronger and more frequent pulsations. Therefore, the structure size appears to be a good predictor for the level of lasing stability. 
	
	The question is what determines the local structure size of lasing modes. The granular structure of lasing modes is formed by the interference of waves propagating in different directions. Thus, the distribution of their directions plays a significant role, which can be unraveled by the spatial Fourier transform of the field profiles. Since it is the \textit{local} directionality that determines the structure size of the intensity distributions, we determine the wave propagation directions in small regions by performing a wavelet transform~\cite{Daubechies1992}, which can be considered as a local Fourier transform [green solid lines in Fig.~\ref{fig4}(c)] \cite{SM}. The local directionality is defined as the inverse of the angular spread of wavelet distributions [Fig.~\ref{fig4}(a)]. The D-cavities and stadia [Figs.~\ref{fig4}(c$_1$-c$_2$)] show wave propagation along almost all directions. The interference of these wave components yields a small structure size. For the \Lim with WGMs [Fig.~\ref{fig4}(c$_3$)], the distribution of wave propagation is more directional parallel to the cavity boundary, leading to a larger structure size perpendicular to the boundary. The ellipses [Fig.~\ref{fig4}(c$_4$)] exhibit a smooth and highly directional distribution, which explains the significantly elongated fine structure of their intensity distributions. Lastly, the square [Fig.~\ref{fig4}(c$_5$)] features four double-peaked lobes along the diagonals, which correspond to the eight plane-wave components of high-Q modes~\cite{Bittner2013b, Bittner2014a}. Despite the low number of plane-wave components, their interference produces nearly isotropic, wavelength-scale intensity grains because the propagation directions are roughly orthogonal.   
	
	To quantify how directional the wave propagation is in a local area, we compute the local directionality $D_W(\vec{r})$ as the inverse of the standard deviation of the wavelet distribution squared \cite{SM}. Figure~\ref{fig4}(d) shows the spatially resolved directionality $D_W(\vec{r})$, averaged over the lasing modes. The D-shaped and stadium cavities [Figs.~\ref{fig4}(d$_1$,d$_2$)] have lower $D_W(\vec{r})$ than \Lim resonators [Fig.~\ref{fig4}(d$_3$)], which demonstrates the difference between spatially extended and whispering-gallery modes. The local directionality of the ellipses [Fig.~\ref{fig4}(d$_4$)] is even higher than for the \Lim, probably due to the integrable ray dynamics of the ellipse, which limits the propagation directions of whispering-gallery trajectories more than for the chaotic trajectories of the \Lim. The squares [Fig.~\ref{fig4}(d$_5$)] have low and almost uniform $D_W(\vec{r})$ over the cavity area.
	
	These different degrees of local directionality originate from the classical ray dynamics. We hence perform ray tracing simulations for cavities with smooth boundaries and define the local directionality $D_R(\vec{r})$ analogously by sampling the long-lived ray trajectories \cite{SM}. Figure~\ref{fig4}(e) shows an excellent agreement between the ray and wave simulations. We average the local directionality over the entire cavity area weighted by the ray intensity, and the results are summarized in Fig.~\ref{fig3}(b) (right axis). The good agreement of local directionality between ray $\langle D_R \rangle$ and wave simulations $\langle D_W \rangle$ indicates that a cavity with sufficiently small boundary roughness can be efficiently simulated by the ray tracing of smooth cavities. More importantly, the strong correlation between $\langle D_R \rangle$ and $\langle s \rangle$ confirms that the structure sizes of lasing modes are determined by the local directionality of optical propagation. Hence, our results demonstrate that ray dynamics can be an efficient tool to qualitatively predict the spatio-temporal lasing dynamics.
	
	Apart from the structure size of lasing modes, spatial localization of the modes can also promote the nonlinear processes in the gain medium. Cavities with WGMs like \Lim or ellipse feature high local optical intensities, and this may facilitate the self-focusing effect and result in stronger output intensity pulsations. Even though simulations with a detailed model of semiconductor carrier dynamics for asymmetric cavities are desirable for a full understanding, our analysis of experiments and passive cavity modes already yields important insights into the relation between the spatial structure of cavity modes and nonlinear lasing dynamics. 
	
	To conclude, we establish the resonator geometry as a powerful design parameter to control the spatio-temporal dynamics of semiconductor microlasers. The lasing dynamics is related to the local directionality of wave propagation, which directly corresponds to the ray dynamics. Our findings enable us to engineer the lasing dynamics by designing the cavity shape based on ray-dynamical principles. In contrast to the design of chaotic microlasers by tailoring temporal oscillation frequencies~\cite{li2022random,Ma2022a}, our approach is based on tailoring the spatial frequencies of lasing modes, which provides a huge and unexplored parameter space. From a practical perspective, customizing the cavity shape enables compact devices to be easily integrated on-chip, in contrast to optical injection and time-delayed feedback~\cite{OhtsuboBook2013, Soriano2013, Sciamanna2015}. Potential applications are the development of high-power broad-area lasers with stable dynamics and compact lasers for chaos-based applications~\cite{Qi2011,Sciamanna2015}. Furthermore, the principle of controlling the nonlinear dynamics via its geometry can also find application in other types of lasers such as broad-area vertical-cavity surface-emitting lasers~\cite{Brejnak2021, bittner2022complex, alkhazragi2023modifying} or random lasers~\cite{bittner2019random}, as well as in other nonlinear dynamic systems in aerodynamics, fluid dynamics, and plasma physics.

	\section*{acknowledgments}
	The authors thank Roland Ketzmerick, Jan Wiersig, Ortwin Hess, Stefano Guazzotti, Takahisa Harayama, and Douglas Stone for fruitful discussions. H. C. and K. K. acknowledge the computational resources provided by the Yale High Performance Computing Cluster (Yale HPC). The work done at Yale is supported partly by the National Science Foundation under GrantNo. ECCS-1953959 and the Office of Naval Research under Grant No. N00014-221-1-2026. S. B. acknowledges funding for the Chair in Photonics by Ministère d'Enseignement Supérieur et de la Recherche (France); GDI Simulation; Re´gion Grand-Est; De´partement Moselle; European Regional Development Fund (ERDF); CentraleSupe´lec; Fondation CentraleSupe´lec; and Metz Metropole. Q. J. Wang, Y. J., and Y. Z. acknowledge National Research Foundation Competi-tive Research Program (NRF-CRP19-2017-01) and National Medical Research Council (NMRC) MOH-000927.

	$^*$ These authors contributed equally.
 
	$^\dagger$ hui.cao@yale.edu

	\clearpage\newpage
	
	\renewcommand{\theequation}{S\arabic{equation}}
	\renewcommand{\thefigure}{S\arabic{figure}}
	\renewcommand{\thetable}{S\arabic{table}}
	\setcounter{figure}{0}
	\setcounter{equation}{0}
	\setcounter{section}{0}

        \section*{Supplemental material}
        
	\subsection{Laser characterization}
	
	\subsubsection{Device fabrication}
	
	We fabricate edge-emitting semiconductor microcavity lasers on a commercial epiwafer with a GaAs/AlGaAs quantum well (Q-Photonics QEWLD-808). First, the bottom metal contact is deposited on the backside of the wafer. Next, lateral boundaries of different cavity shapes are defined via UV-lithography and inductively coupled plasma (ICP) dry etching. The etch depth is 3.5 $\mu$m, reaching the bottom cladding layer to ensure strong light confinement within the cavity by a high refractive index difference at the sidewalls. The sidewalls are almost vertical and have a small but non-negligible surface roughness (see Ref.~\cite{Cerjan2019}). Finally, the top metal (Ti/Au) contacts for electrical pumping are deposited. The top contacts are withdrawn by 6 $\mu$m from the cavity edges in order to avoid misalignment during photolithography, causing the metal hanging down over the sidewall and blocking the edge emission (see Ref.~\cite{Kim2022} for details of the fabrication process). 
	
	\begin{figure*}[t]
		\begin{center}
			\includegraphics[width = 17 cm]{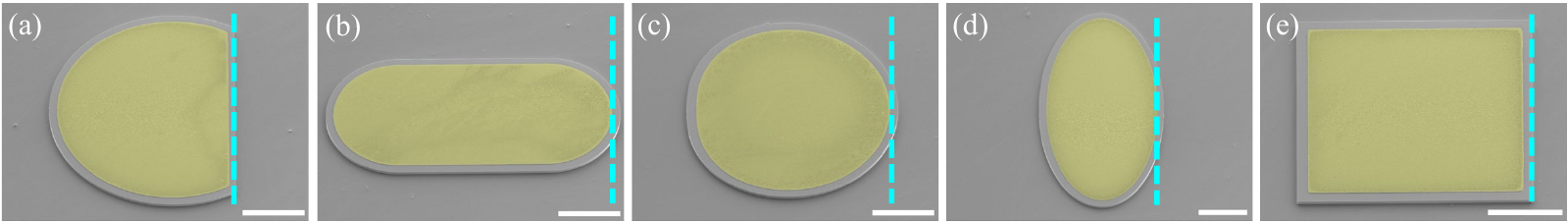}
		\end{center}
		\caption{Scanning electron microscope images of GaAs microlasers with D-shaped (a), stadium (b), \Lim (c), ellipse (d), and square (e) dielectric cavities in perspective view. The scale bars are $50~\mu$m long. The yellow area represents the top metal contact for electrical pumping. The cyan dashed lines mark the facets imaged onto the streak camera.}
		\label{fig:SEM}
	\end{figure*}
 
	Multiple devices for each of the five different geometries (D-shaped, stadium, \Lim, ellipse, and square) are fabricated concurrently on the same epiwafer (see Fig.~\ref{fig:SEM}). The D-cavity is a circle with a straight cut, where the radius is $R_D = 100~\mu$m, and the distance from the circle center to the cut is set to $R_D/2$ to render the ray dynamics as chaotic as possible. The stadium consists of a square with side length $a = 119~\mu$m between two semicircles of radius $a/2$. The boundary of the \Lim cavity \cite{Robnik1983,Wiersig2008,kim2023spatiotemporal} is defined in polar coordinates by $\rho(\varphi) = \rho_0 (1 + \epsilon \cos\varphi)$ where $\varphi$ is the azimuthal angle, $\rho_0 = 86~\mu$m the mean radius and $\epsilon = 0.42$ the deformation parameter which is chosen to obtain unidirectional emission~\cite{Wiersig2008}. The ellipse cavities have an aspect ratio of $b/a = 2$ with a minor (major) diameter of $a = 127~\mu$m ($b = 254~\mu$m). Lastly, the square microlasers have a side length of $159~\mu$m. These different cavity geometries result in approximately equal cavity areas ($2.53\times10^4~\mu\mathrm{m}^2$).
	
	The guiding layer of the quantum well wafer supports a single vertical excitation mode, and thus the optical field propagates in the plane of the resonators with a phase velocity $c / \neff$, where $c$ is the speed of light in vacuum and $\neff = 3.37$ the effective refractive index. The laser emission has transverse electric (TE) polarization; that is, the electric field is parallel to the plane of the resonators. 

	\subsubsection{Device testing}
	
	We perform experiments on five microlasers for each cavity shape. The microlasers are contacted with a Tungsten needle (Quater Research H-20242) and pumped electrically at room temperature by a diode driver (DEI Scientific PCX-7401). We use  $2~\mu$s long current pulses with a repetition rate lower than 1 Hz to reduce heating. All the spatio-temporal measurements are  done at the same current of $500$~mA (corresponding to about $2.0~$kA/cm$^2$) for all cavity shapes. 
	
	The optical spectra of lasing emission, with a central wavelength of about 800 nm, are measured with an imaging spectrometer (Acton SP300i) equipped with an intensified CCD camera (ICCD, Andor iStar DH312T-18U-73). For measurements of time-resolved optical spectra, we time-gate the ICCD camera with a temporal resolution of 50 ns.
	
	To investigate the spatio-temporal dynamics of the microlasers, one facet of a microcavity (marked by cyan dashed lines in Fig.~\ref{fig:SEM}) is imaged onto the entrance slit of a streak camera using a microscope objective ($20\times$, $\mathrm{NA}=0.4$) and a tube lens. Then the near-field laser emission is measured by a streak camera (Hamamatsu C5680) with a fast single-sweep unit (M5676) (see Ref.~\cite{Bittner2018a} for further details of the setup). We measure 10 ns long streak images with a temporal resolution of about 30 ps.

	\subsubsection{Lasing thresholds}
	
	\begin{figure}[b]
		\begin{center}
			\includegraphics[width = 8.5 cm]{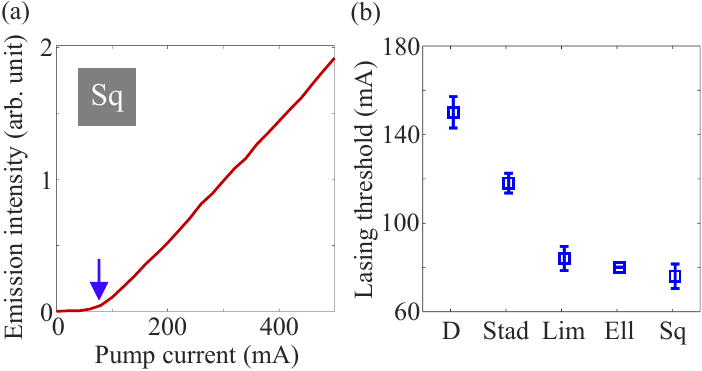}
		\end{center}
		\caption{Experimentally measured lasing thresholds of asymmetric cavities with deformed geometry. (a) An exemplary LI curve of a square microcavity laser: emission power obtained from the time-integrated lasing spectra as a function of injected current. The lasing threshold is 80 mA (blue arrow). (b) Lasing thresholds for the five cavity geometries. The errorbars represent the variation between five different devices of the same geometry.}
		\label{fig:thres}
	\end{figure}
	
	Figure~\ref{fig:thres} shows the lasing thresholds for different resonator geometries. The LI curves of all the measured cavities exhibit a sharp transition in the slope, where one example for a square cavity is shown in Fig.~\ref{fig:thres}(a). Lasing thresholds for different cavity shapes, each averaged over five devices, are presented in Fig.~\ref{fig:thres}(b). The fluctuations of the threshold amongst cavities of the same shape are very small, indicating the generally good reproducibility and consistency of fabrication and measurements. The typical lasing thresholds range from 80 to 150 mA (from 0.32 to 0.60 kA/cm$^2$), strongly depending on the cavity geometry. The square, ellipse, and \Lim cavities have the low thresholds, followed by the stadium, and the D-cavities have the highest thresholds.

	\subsection{Lasing dynamics}
	
	\subsubsection{Optical spectrum} \label{sec:specChrono}

	\begin{figure}[b]
		\begin{center}
			\includegraphics[width = 7.5 cm]{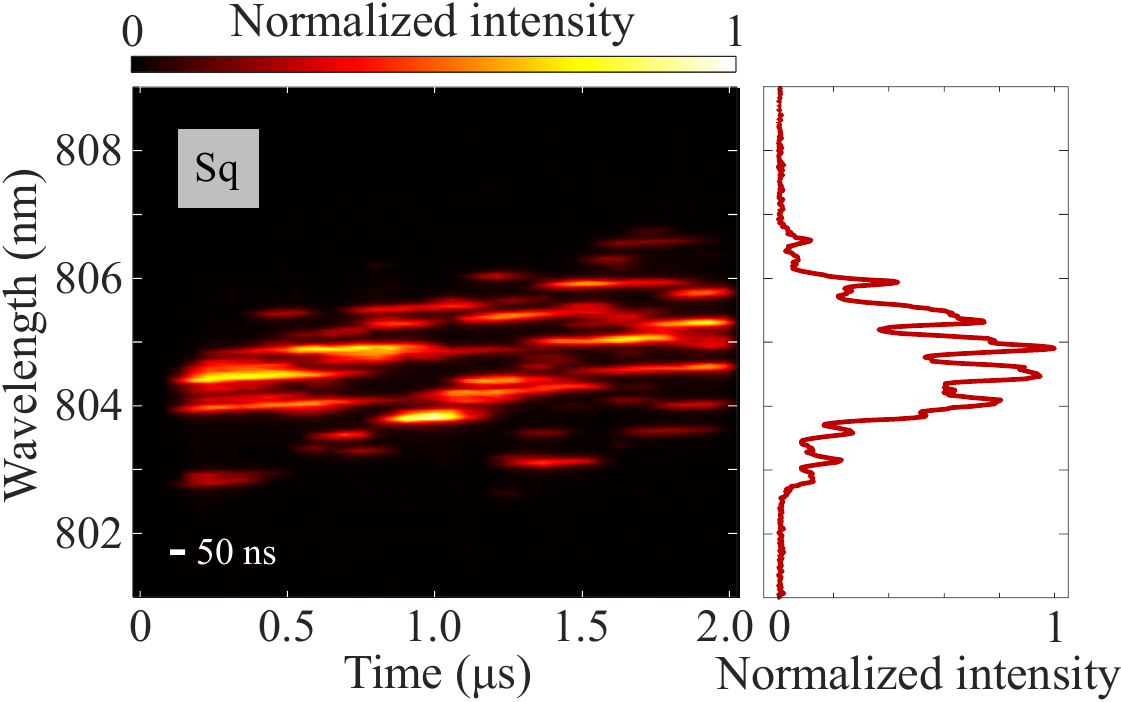}
		\end{center}
		\caption{Measured spectrochronogram of a square microcavity laser at $500$~mA pump current. The horizontal axis indicates the time delay $t_d$ of the gate interval with respect to the start of the 2-$\mu$s-long current pulse. The white scale bar indicates the time resolution of $50$~ns. The right panel shows the emission spectrum integrated over the entire pulse, exhibiting multiple overlapping lasing peaks.}
		\label{fig:specChrono}
	\end{figure}
	
	Figure~\ref{fig:specChrono} shows the spectrochronogram (time-resolved spectrum) of a square microlaser which is measured with a tomographic technique: the ICCD connected to the spectrometer is gated on during a $50$~ns long interval to measure spectra with a high temporal resolution, and the time delay time $t_d$ of the gate interval with respect to the start of the pump pulse is successively increased from one pulse to the next (cf.\ Ref.~\cite{Bittner2018a}). The laser remains multimode at any point in time, though the instantaneous number of lasing peaks is smaller than that in the time-integrated spectra (the right panel of Fig.~\ref{fig:specChrono}). Similar spectrochronograms are obtained for the other microlaser geometries (cf.\ Refs.~\cite{Bittner2018a,Bittner2020,kim2023spatiotemporal}). It should be emphasized that multimode lasing is observed for all cavity shapes, and the interaction of multiple lasing modes is an important ingredient in the development of lasing instabilities \cite{OhtsuboBook2013, Ma2022a}.
	
	The spectrum changes continuously over the course of the current pulse with a timescale of the order of $100$~ns, which is much longer than the intrinsic time scales of the nonlinear lasing dynamics discussed in Section~\ref{sec:RFprocess}. Due to Joule heating, the gain spectrum redshifts, thus the lasing peaks at shorter wavelength disappear, and new peaks emerge at longer wavelengths \cite{Bittner2018a}. Consequently, the lasing state gradually evolves over the course of the pulse. This enables us to investigate an ensemble of different possible dynamical states by recording streak images at different times during the pulse, where the time window (10 ns) of individual streak images is considerably shorter than the time scale of the heating-induced drift of the lasing spectrum. 
	
	\subsubsection{Spatio-temporal dynamics}
	
	We measure the evolution of spatio-temporal lasing dynamics during a pump current pulse of 2 $\mu$s length.  We recorded single-sweep streak images of 10 ns length (temporal resolution about $30$~ps). While repeatedly pumping the lasers under identical conditions, we gradually shift the gated time window of measurement to scan the entire emission pulse. More specifically, the delay time $t_d$ between the streak image and the beginning of the current pulse is varied in the range of 0.3 -- 1.9 $\mu$s \cite{Bittner2018a}. Each streak image provides the spatio-temporal intensity pattern of lasing emission $I(x, t_d, t_d+t)$, where $t = 0$--$10$~ns is the time during one image. For each microlaser, $161$ consecutive images covering a total time of $1.61~\mu$s are measured, omitting only the transient dynamics at the start and end of the 2-$\mu$s-long pulses. These images are concatenated in the time domain to obtain microsecond-long traces. To account for cavity-to-cavity variations, we obtain streak images for five different devices for each cavity geometry. The examples for each cavity geometry are shown in Fig.~2(a) of the main text.
	
	Several dynamic processes in our microlasers contribute to the spatio-temporal fluctuations observed in the streak images \cite{kim2023spatiotemporal}. First, there are relatively strong intensity fluctuations with a rather narrow radio-frequency (RF) band, which appear as peaks in the short-term RF-spectra for some delay times $t_d$ during the current pulses and can be spatially localized as well. Second, we observe a ubiquitous spatio-temporal speckle caused by the beating of many lasing modes with different optical frequencies and spatial patterns \cite{Kim2021, Kim2022} and/or amplified spontaneous emission with a broad spectrum. This spatio-temporal speckle is common for multimode lasers like the ones investigated here but is not the focus of the current study. Hence we will separate the first type of fluctuations, resulting from filamentation and lasing instabilities, from the second type due to spatio-temporal interference. 
	
	\subsubsection{Radio-frequency spectrum} \label{sec:RFprocess}
	
	\begin{figure*}[hbtp]
		\begin{center}
			\includegraphics[width = 13 cm]{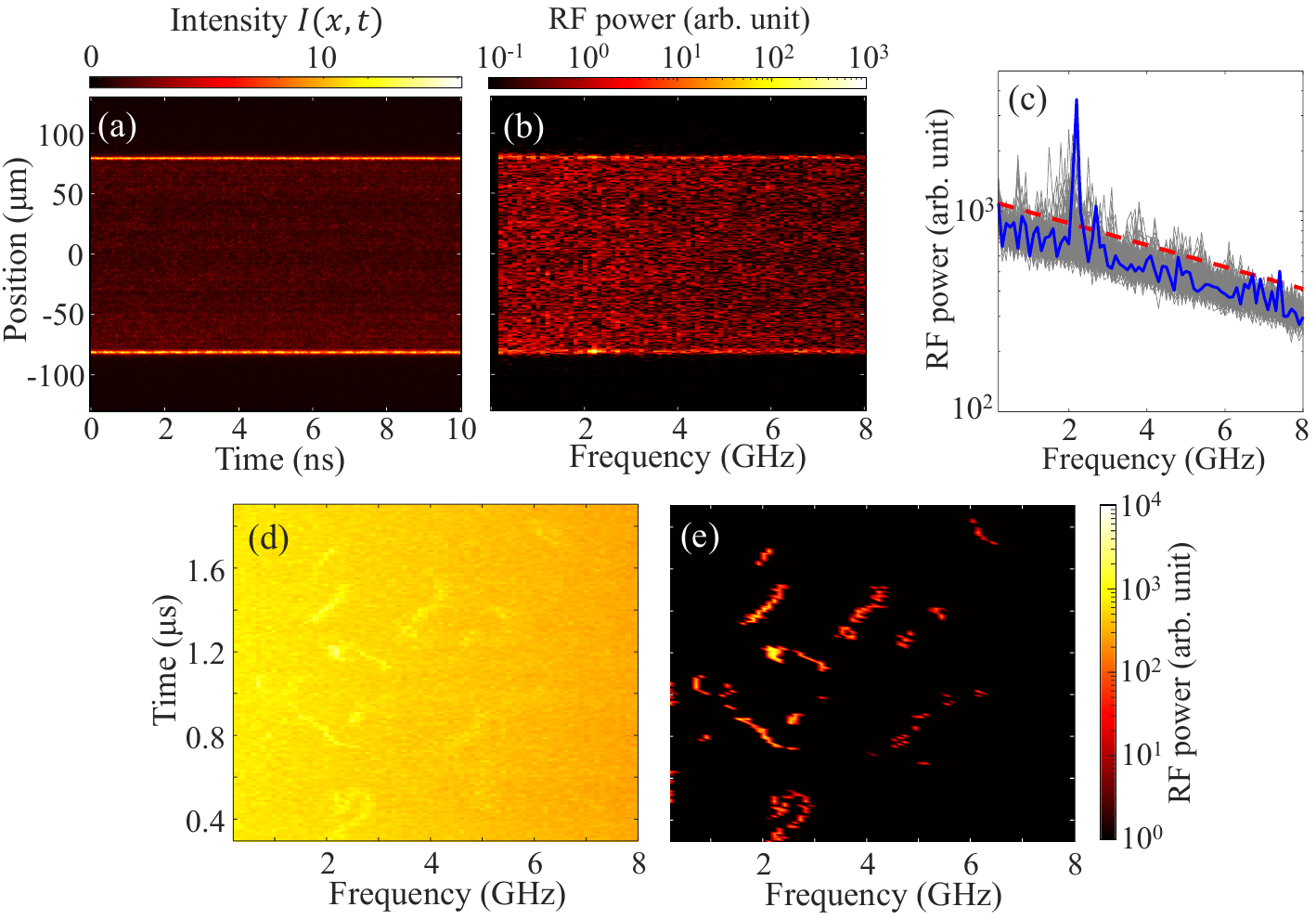}
		\end{center}
		\caption{Subtraction of RF-spectrum to highlight frequency components from lasing instabilities. (a)~Streak image of lasing emission from a square microcavity for a time window of 10 ns length, measured at the delay time $t_d$ = 1.20 $\mu$s from the beginning of a 2-$\mu$s-long pulse. (b)~Spatially resolved RF spectrum $|\tilde{I}(x, f, t_d)|^2$ of the emission intensity, obtained from temporal Fourier transform of (a). (c)~Spatially-averaged RF-spectra $S(f, t_d)$ at all delay times $t_d$ are plotted together (gray curves). The blue curve highlights $S(f, t_d)$ from (b), which exhibits a peak standing out at 2.2 GHz. The red dashed line indicates the fit $S_\mathrm{b}(f)$ of the exponential decay of the RF-power from the resolution-limited spatio-temporal speckle. The RF peaks above the red line result from lasing instabilities. (d)~Time-resolved RF-spectrum $S(f, t_d)$ before subtracting $S_\mathrm{b}(f)$. (e)~Modified RF-spectrum $\hat{S}(f, t_d)$ obtained by subtracting $S_\mathrm{b}(f)$ from $S(f, t_d)$. The peaks due to lasing instabilities are clearly visible once the contributions from spatio-temporal beating are removed.}
		\label{fig:RFprocess}
	\end{figure*}
	
	Here we separate intensity fluctuations of different origins in the measured streak images. First, each streak image $I(x, t_d, t_d+t)$ is normalized such that $\langle I(x, t_d, t_d+t) \rangle_{x, t} = 1$ to enable a quantitative comparison of the fluctuation strength of different measurements. Next, we calculate the spatially-resolved Fourier transform (FT) of the streak image,
	\begin{equation} \tilde{I}(x, f, t_d) = \int_0^{T} dt \, I(x, t_d, t_d+t) e^{-2 \pi i f t} \, , \end{equation}
	where $f$ is the frequency in the range up to $12$~GHz, which is limited by the temporal resolution of the streak image, and $T$ = 10 ns is the length of a single streak image. Then we spatially average it to obtain the RF power spectrum
	\begin{equation} S(f, t_d) = \langle |\tilde{I}(x, f, t_d)|^2 \rangle_x \end{equation}
	of the streak image. 
	
	The spatio-temporal beating of lasing modes and/or ASE produces a very broad range of RF components. The highest possible frequency, given by the width of the optical spectrum of lasing emission, is of the order of hundreds of GHz. Hence, the measured spatio-temporal speckle is limited by the temporal resolution of our streak camera. The Lorentzian-shaped point spread function for the temporal response of our streak camera \cite{Kim2021} results in a continuous RF spectrum that decays exponentially with frequency. The lasing instabilities manifest in relatively narrow peaks on top of it. 
	
	We develop a process to separate the RF peaks from the broadband contribution of spatio-temporal speckle. Figure~\ref{fig:RFprocess}(a) shows an exemplary streak image of a square microcavity laser, which contains intensity fluctuations from lasing instabilities as well as spatio-temporal beating. Temporal Fourier transform gives the spatially-resolved RF spectrum in Fig.~\ref{fig:RFprocess}(b). Since the RF power of the spatio-temporal speckle has a comparable magnitude to that of the lasing pulsations, it is not easy to visually separate the two. The spatially-averaged RF-spectrum [blue curve in Fig.~\ref{fig:RFprocess}(c)] displays a peak at frequency $2.2$~GHz due to almost regular intensity oscillations.
 
    To determine the RF-power level of the spatio-temporal speckle, we plot the spatially-averaged RF-spectra $S(f, t_d)$ at every time $t_d$ together as shown in Fig.~\ref{fig:RFprocess}(c). All spectra feature a continuous signal decaying exponentially with frequency, which originates from the resolution-limited spatio-temporal speckle \cite{Kim2021}. We perform a linear fit of the exponential decay in a semi-logarithmic scale. It yields the broadband continuous RF-signal $S_{\mathrm{b}}(f)$, which we consider as the RF-power level of spatio-temporal speckle.
	
	To separate the RF peaks above $S_{\mathrm{b}}(f)$, we subtract this fitted exponentially-decaying power spectrum by
	\begin{equation} \hat{S}(f, t_d) = S(f, t_d) - S_{\mathrm{b}}(f), \end{equation}
	and setting the negative values of $\hat{S}(f, t_d)$ to 0. The subtracted RF-spectrum $\hat{S}(f, t_d)$ is presented in Fig.~\ref{fig:RFprocess}(e), in which only the oscillatory peaks caused by lasing instabilities are left and can thus be well discerned. Figure~2(b) in the main text shows the subtracted RF-spectra $\hat{S}(f, t_d)$ for each cavity geometry.

	\subsubsection{Lasing instabilities}
	
	For a more quantitative evaluation of the RF-spectra, we consider two quantities characterizing different aspects in the following. First, we look at the integrated RF power,
	\begin{equation} S_\mathrm{tot} = \sum \limits_{f, t_d} \hat{S}(f, t_d) \, . \end{equation}
	Second, we calculate the frequency of occurrence of instabilities via the participation ratio of the RF spectra,
	\begin{equation} S_{\mathrm{PR}} = \frac{\langle \hat{S}(f, t_d) \rangle^2_{f, t_d}}{\langle [\hat{S}(f, t_d)]^2 \rangle_{f, t_d}} \, . \end{equation}
	These two quantities are extracted from measurements with five different microlasers per geometry, and the results averaged in logarithmic scale are presented in Fig.~3(a) of the main text.

	\subsection{Wave simulations} \label{sec:sim}
	
	\subsubsection{Resonances}
	
	We model the laser cavities as two-dimensional (2D) dielectric resonators, that is, the passive cavity modes are the solutions of the 2D scalar Helmholtz equation
	\begin{equation} \{ \Delta + n^2(\vec{r}) k^2 \} \Psi(\vec{r}) = 0 \, , \label{eq:helmholtz} \end{equation}
	with outgoing-wave boundary conditions \cite{Tureci2005, Cao2015}, where $\vec{r} = (x, y)$ are the coordinates in the plane of the cavity, $n(\vec{r})$ is the effective refractive index structure of the cavity, $n(\vec{r}) = \neff=3.37$ inside and $n(\vec{r}) = 1$ outside of the cavity, and $\Psi$ corresponds to the $z$-component of the magnetic field, $H_z$, for TE-polarized modes. The solutions are computed numerically using the COMSOL eigenfrequency solver module, where the outgoing-wave boundary conditions are implemented via perfectly matched layers at the boundaries of the computational domain. A real part of the resonance wave numbers $k$ corresponds to the resonance frequency and the imaginary part to the decay rate (inverse of the lifetime), hence the quality factors are given by $Q = -\Re(k) / [2 \Im(k)]$. 
	
	The wave simulations are performed for cavities with ten times smaller linear dimensions (area of about $253~\mu\mathrm{m}^2$) due to computational constraints. Since both the simulated and the fabricated cavities are well within the semiclassical regime $kR \gg 1$, where $R$ is a typical linear dimension (e.g., the radius of the D-cavity), the simulation results of smaller cavities can nonetheless be considered representative of the actual, larger cavities. Exemplary high-Q resonances of five different cavity geometries with smooth boundaries are shown in Fig.~1(b$_1$-b$_5$) of the main text.
	
	\subsubsection{Surface roughness} \label{sec:rough}
	
	Since surface roughness is inevitable in fabricated microresonators, we calculate the resonances of passive resonators with boundary roughness. The roughness model we use is similar to the one in Refs.~\cite{liew2015pump, Bittner2018a}. The cavity boundaries are perturbed by a superposition of high-order harmonics with random phase and amplitude,
	\begin{equation}
	\label{roughDef} 
	\Delta R(s) = C\sum^{m_2}_{m=m_1} a_m \mathrm{cos}(2\pi ms+\theta_m),
	\end{equation}
	where $s\in [0,1)$ is the normalized arc length coordinate along the cavity boundary. At every location $s$, the perturbation is added in the direction perpendicular to the boundary. The random variables $a_m$ determine the amplitude of harmonic perturbation, which follow a uniform distribution in the range of $a_m \in [-A, A]$~nm, where $A$ is set to 10 nm. The random phases $\theta_m$ are uniformly distributed in the range $[0,2\pi)$. The normalization constant $C$ is given by $\sqrt{6/(m_2-m_1+1)}$ so the root-mean-square of the fluctuation $\Delta R(s)$ is $A$.
	
	The lowest-order harmonics $m_1$ (with the longest period) is determined by $s_0 / L_{\mathrm{max}}$, where $s_0$ is the perimeter of the cavity boundary and $L_{\mathrm{max}}$ is the maximal length scale of boundary fluctuations parallel to the cavity boundary. Based on high-resolution SEM images of experimental cavities, we set $L_{\mathrm{max}} = 1~\mu$m. The highest harmonics $m_2$ (with the shortest period) is determined by $s_0 / L_{\mathrm{min}}$, where $L_{\mathrm{min}}$ is the shortest scale of boundary modulation considered. We set its value to half of the in-medium wavelength, $L_{\mathrm{min}} = \lambda / (2 \neff) = 0.12~\mu$m, as any perturbation on length scales much finer than the wavelength cannot be resolved by light. Given that the perimeters $s_0$ of the simulated cavities of all five different shapes are roughly 60 $\mu$m, the approximate range of the harmonics $m$ is from $m_1$ = 60 to $m_2$ = 500, where the exact values of $m_1$ and $m_2$ depend on the cavity shape.

	\subsubsection{Lasing modes} 
	
	An important aspect of microcavity lasers is the mode competition that determines the number of lasing modes. While simulating the dynamics of our asymmetric microcavity lasers is extremely computationally expensive \cite{Bittner2018a}, the Steady-state \textit{ab-initio} Lasing Theory (SALT) \cite{Ge2010, Cerjan2016} has proven useful to estimate the number of lasing modes of such lasers, although it is strictly speaking applicable only to lasers that have reached a steady-state. As our microcavities have relatively high Q-factors, we employ the single-pole approximation to SALT (SPA-SALT), which assumes that each lasing mode is represented by a single passive cavity mode~(see Refs.~\cite{Ge2010, liew2015pump, Cerjan2016, Cerjan2019} for details). 
	
\begin{figure}[hbtp]
		\begin{center}
			\includegraphics[width = 8 cm]{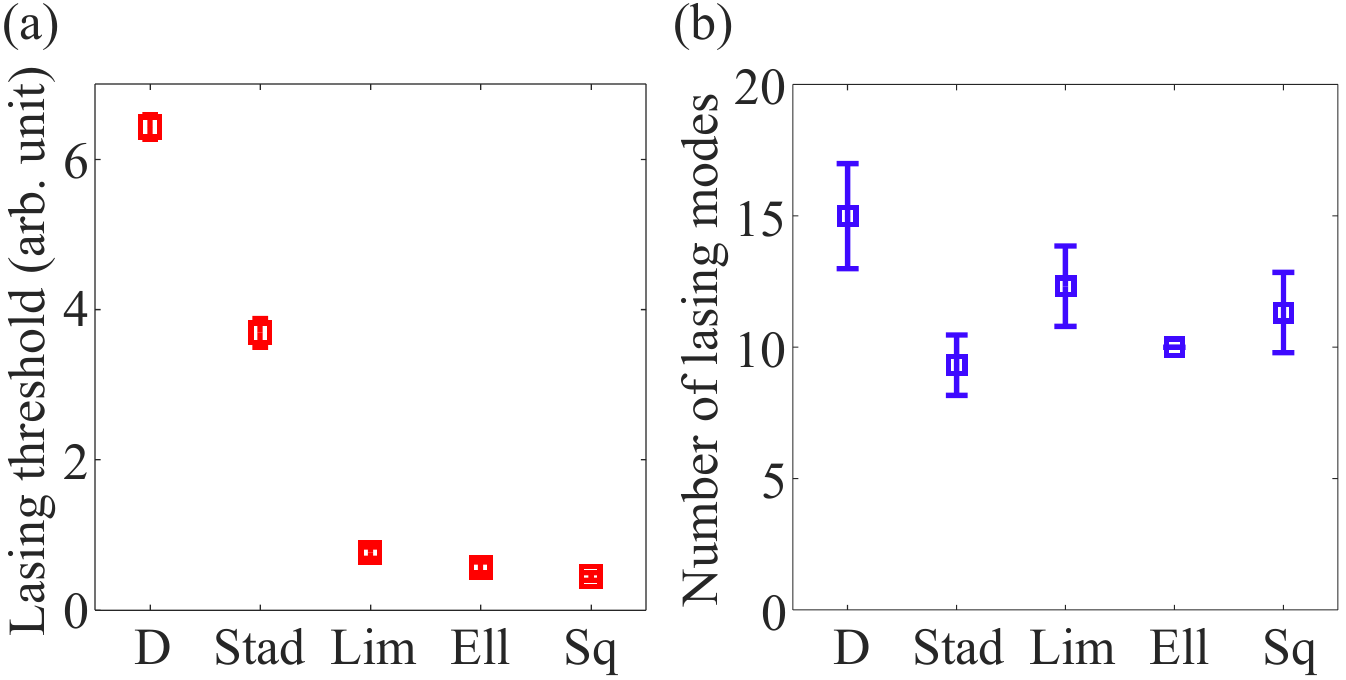}
		\end{center}
		\caption{Lasing behaviors simulated by SPA-SALT. (a) Lasing thresholds and (b) the number of lasing modes at ten times the lasing threshold for five different cavity geometries. The errorbars denote the variation from three different realizations of surface roughness for each geometry.}
		\label{fig:SALT}
	\end{figure}

	We perform SPA-SALT simulations for three different realizations of rough boundaries for each cavity shape. All high-Q resonances within the spectral range of $797$--$803$ nm are considered. The first lasing threshold is inversely proportional to the quality factor $\Qmax$ of the most long-lived mode. Figure~\ref{fig:SALT}(a) shows the computed lasing thresholds for the five cavity shapes. \Lim, ellipse, and square cavities have the lowest thresholds, whereas those of D-cavity and stadium are significantly higher. This trend is consistent with the experimental data [Fig.~\ref{fig:thres}(b)], and demonstrates how strongly the cavity shape affects the lasing threshold.	
	
	The onset of further lasing modes is determined by a combination of their $Q$-factors compared to the first lasing mode, which saturates the gain, and the spatial overlap of their intensity distributions. The actual number of lasing modes and the powers of lasing modes depend on spatial gain-competition effects, which are taken into account by the SPA-SALT simulations. We perform SPA-SALT simulations at $10$ times the lasing threshold even though we pump less strongly in experiments. This is because the microcavities in simulation have much smaller sizes and hence a smaller number of lasing modes than those in experiments. Figure~\ref{fig:SALT}(b) shows the computed number of lasing modes for the five cavity shapes. All cavity shapes exhibit multimode lasing, with a slightly different number of lasing modes at 10 times their lasing thresholds. Note that individual lasing modes have different powers, given by the SPA-SALT simulations.
	
	\begin{figure*}[hbtp]
		\begin{center}
			\includegraphics[width = 17 cm]{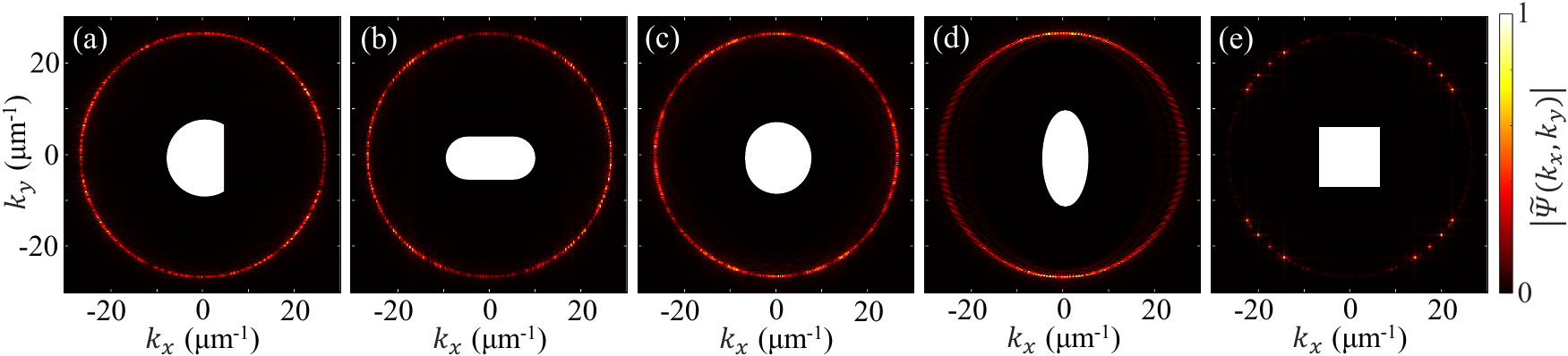}
		\end{center}
		\caption{Spatial Fourier transform of resonances in (a)~D-cavity, (b)~stadium, (c)~\Lim, (d)~ellipse and (e)~square microcavities with surface roughness. The spatial distributions of field amplitude of the modes are shown in Figs.~4(b$_1$-b$_5$) of the main text. The chaotic cavities (a-c) have uniformly distributed plane-wave components in all directions, while ellipse (d) and square (e) resonators feature dominant propagation in vertical and diagonal directions, respectively. The Fourier transform, however, does not include information about spatially-resolved propagation directions.}
		\label{fig:spatialFT}
	\end{figure*}
	
	\subsubsection{Spatial Fourier transform} \label{sec:spatialFT}
	
	The main objective of wave simulations is to investigate how the cavity shape affects the optical propagation directions inside the cavity. A straightforward approach is performing the Fourier transform of a passive cavity resonance. The spatial Fourier transform of its field distribution $\Psi(\vec{r})$ is
	\begin{equation} \tilde{\Psi}(\vec{k}_M) = \int d\vec{r} \, \Psi(\vec{r}) \, \exp[-i (\vec{k}_M \cdot \vec{r})] \, , \end{equation}
	where $\vec{k}_M = (k_x, k_y)$ is the in-medium wave vector, and the integral is over the interior of the cavity, corresponding to an expansion of the field distribution in plane-wave components.
	
	The wave-vector distributions of the resonances shown in Figs.~4(b$_1$-b$_5$) of the main text are presented in Fig.~\ref{fig:spatialFT}. In general, all plane-wave components have the same $|\vec{k}_M| = 2 \pi n / \lambda$, where $n$ is the refractive index and $\lambda$ is the resonant wavelength, and hence all the wave-vector distributions in Fig.~\ref{fig:spatialFT} are localized on a circle with that radius. For the D-cavity, stadium, and \Lim resonators [Figs.~\ref{fig:spatialFT}(a-c)], the circle is covered quite homogeneously, indicating that there is no preferred propagation direction of the underlying ray trajectories. This is not surprising in view of their chaotic ray dynamics [cf.\ Figs.~1(a$_1$-a$_3$)]. For the ellipse [Fig.~\ref{fig:spatialFT}(d)], the components propagating in the vertical direction (parallel to the major axis) are stronger. This is due to the vertically elongated shape of the ellipse: the whispering gallery trajectories spend more time propagating parallel to the approximately vertical boundaries since these are longer compared to the approximately horizontal boundary parts near the vertices [see Fig.~4(a$_4$)]. For the square [Fig.~\ref{fig:spatialFT}(e)], the wave-vector components are stronger in the diagonal directions $\theta = \pm 45^\circ$ and $\pm 135^\circ$ since trajectories traveling in these directions are most strongly confined by TIR. For a smooth square, the high-$Q$ modes consist of exactly $8$ plane-wave components with directions close to the diagonals \cite{Bittner2013b, Bittner2014a}, but the surface roughness creates additional propagation directions via scattering and thus forms the four clusters of peaks around the diagonals that we observe in Fig.~\ref{fig:spatialFT}(e). 
	
	The main drawback of the spatial Fourier transform, in particular for analyzing resonances of wave-chaotic cavities, is that the information about the propagation directions comes without spatial resolution. Hence, in order to obtain the propagation directions locally in the cavity, we apply a wavelet transform. 

	\subsection{Local directionality and structure size} \label{sec:locdirfeat}
	
	\subsubsection{Wavelet transform} \label{sec:wavelet}
	
	\begin{figure}[hbtp]
		\begin{center}
			\includegraphics[width = 8 cm]{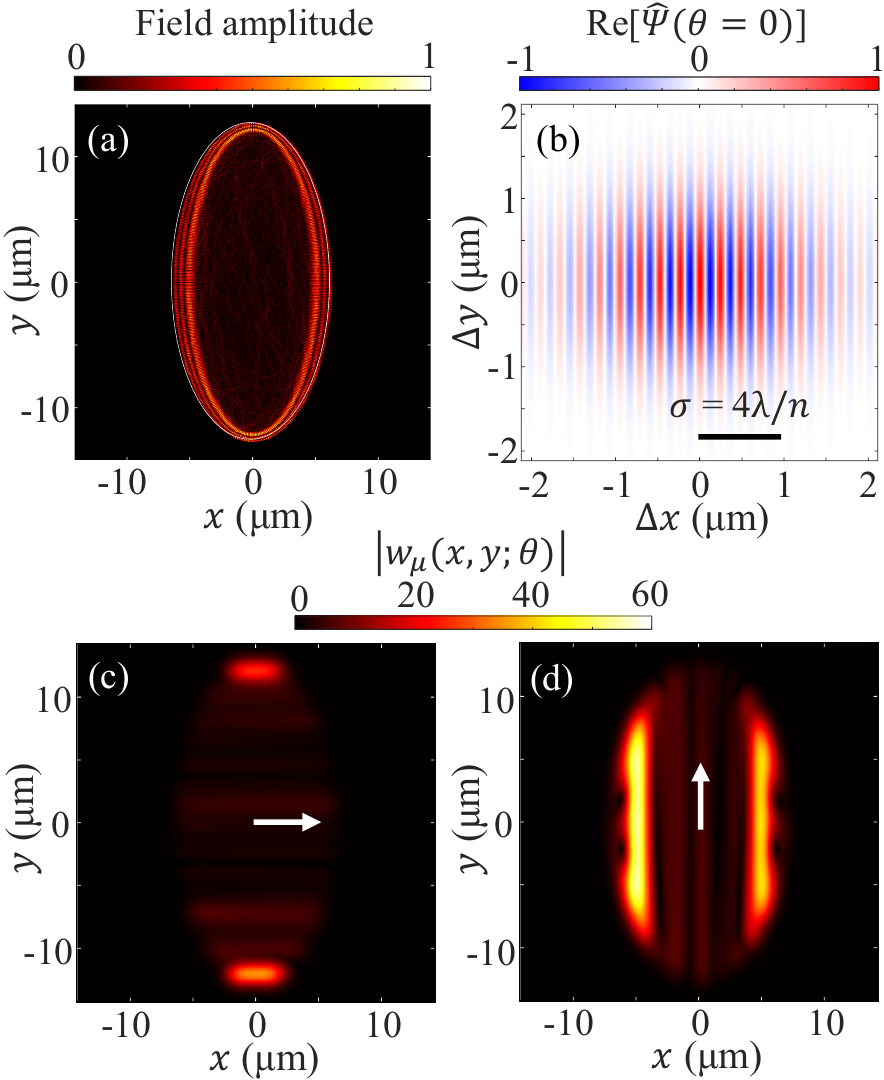}
		\end{center}
		\caption{Wavelet transform of a cavity resonance. (a) Field amplitude of a high-Q mode ($\lambda = 797.5$~nm, $Q = 1.56\times10^4$) in an ellipse resonator with rough boundary. (b) Real part of the Morlet wavelet $\hat{\Psi}$ with orientation $\theta = 0^\circ$. Its standard deviation $\sigma$ is 0.95 $\mu$m, four times the in-medium wavelength. The imaginary part of $\hat{\Psi}$ is similar to its real part, except for a phase shift of $\pi/2$. (c)~Wavelet transform $|w_\mu(\vec{r}, \theta)|$ of the wave function for $\theta = 0^\circ$ and (d)~for $\theta = 90^\circ$, where the orientation is indicated by a white arrow. The wavelet transform reveals the dominant propagation directions with spatial resolution.}
		\label{fig:wavelet}
	\end{figure}
	
	We calculate the wavelet transform of spatial field distribution $\Psi_\mu$ of the lasing mode $\mu$, which can be considered a local Fourier transform \cite{Daubechies1992}, as
	\begin{equation} w_\mu(\vec{r}, \theta) = \int d\vec{r}\,' \, \Psi_\mu^*(\vec{r}\,') \hat{\Psi}(\vec{r}\,' - \vec{r}, \theta) \, , \end{equation}
	where $\hat{\Psi}$ is the Morlet wavelet [Fig.~\ref{fig:wavelet}(b)]
	\begin{equation} \hat{\Psi}(\vec{r}, \theta) = e^{-i k_M x_R -x_R^2 / (2 \sigma^2) -y_R^2 / \sigma^2} \, . \end{equation}
	The rotated coordinate frame $(x_R, y_R)$ is given by
	\begin{equation} \begin{array}{rcl} x_R & = & x \cos\theta + y \sin\theta \\ y_R & = & -x \sin\theta + y \cos\theta \end{array} \end{equation}
	where $\theta$ is the azimuthal angle with respect to the $x$-axis and $k_M = 2 \pi n / \lambda$ is the in-medium wave number of the mode $\Psi_\mu$. The width $\sigma$ of the Morlet wavelet is $\sigma = 4\lambda/n = 0.95~\mu$m, which is four times the in-medium wavelength of resonant modes. Therefore, $|w_\mu(\vec{r}, \theta)|$ yields the amplitude of waves propagating in direction $\theta$ in a small region of approximate diameter $2 \sigma$ around $\vec{r}$. 
	
	Figure~\ref{fig:wavelet} exemplifies the wavelet transform using a wave function of an ellipse resonator with surface roughness. The field amplitude is shown in Fig.~\ref{fig:wavelet}(a) and the Morlet wavelet for $\theta = 0^\circ$ in Fig.~\ref{fig:wavelet}(b). The wavelet transform for $\theta = 0^\circ$, that is, horizontal wave propagation, is shown in Fig.~\ref{fig:wavelet}(c). It features high amplitudes only close to the two vertices of the ellipse where the whispering gallery trajectories propagate horizontally [cf.\ Fig.~1(a$_4$)]. In contrast, the wavelet transform for vertical propagation ($\theta = 90^\circ$) in Fig.~\ref{fig:wavelet}(d) features high intensity near the vertically extended parts of the boundaries where ray trajectories propagate in this direction. This demonstrates how the wavelet transform gives spatially resolved information of the wave propagation directions, which corresponds well to the classical dynamics, in contrast to the spatial Fourier transform shown in Fig.~\ref{fig:spatialFT}(d). 

	\subsubsection{Local directionality}
	
	We calculate the spatially-resolved local directionality of wave propagation over the entire cavity area. First, we calculate the integrated wavelet distribution for all lasing modes $\mu$,
	\begin{equation}
	W(\vec{r},\theta) = \sum_{\mu} P_{\mu}|w_\mu(\vec{r},\theta)|^2,
	\end{equation}
	and the average intensity profile
	\begin{equation}
	I(\vec{r}) = \sum_{\mu} P_{\mu} \int |w_\mu(\vec{r},\theta)|^2 d\theta.
	\label{eq:intensity}
	\end{equation}
	Here $P_{\mu}$ is the lasing mode power from SPA-SALT, so the lasing modes with higher power contribute more to the average. The local angular spread of propagation is then defined as
	\begin{equation}
	\sigma_W(\vec{r})=\min_{\theta_0}\sqrt{ \frac{ \int^{+\frac{\pi}{2}}_{-\frac{\pi}{2}}\theta^2[W(\vec{r},\theta+\theta_0) + W(\vec{r},\theta+\theta_0-\pi)]d\theta }{ \int^{+\pi}_{-\pi}W(\vec{r},\theta)d\theta } }.
	\label{eq:angWave}
	\end{equation}
	Here $\theta_0$ indicates the orientation to calculate the angular spread, and the two terms in the numerator indicate the forward and backward half of the angular domain with respect to $\theta_0$. We search for the direction $\theta_0$ in which the angular spread is minimal. It corresponds to the dominant propagation direction. Equation~(\ref{eq:angWave}) yields the minimal angular spread in terms of the standard deviation of the wavelet distribution as a function of the polar angle, and the local directionality is defined as its inverse,
	\begin{equation}
	D_W(\vec{r}) = \frac{1}{\sigma_W(\vec{r})}.
	\end{equation}
	Maps of the local directionality $D_W(\vec{r})$ for cavities with rough boundaries and different geometries are presented in Figs.~4(d$_1$-d$_5$) of the main text.
	
	Finally, we calculate the overall local directionality of a cavity by averaging $D_W(\vec{r})$ over the position weighted by the local intensity,
	\begin{equation} 
	\langle D_W \rangle = \frac{\int d\vec{r} \, D_W(\vec{r}) I(\vec{r}) }{ \int d\vec{r} \, I(\vec{r}) } \, . \end{equation}
	The mean local directionality $\langle D_W \rangle$ for different cavities is presented in Fig.~3(b) of the main text.

	\subsubsection{Local structure size}
	\label{sec:structSize}
	
	\begin{figure}[ht]
		\begin{center}
			\includegraphics[width = 8.5 cm]{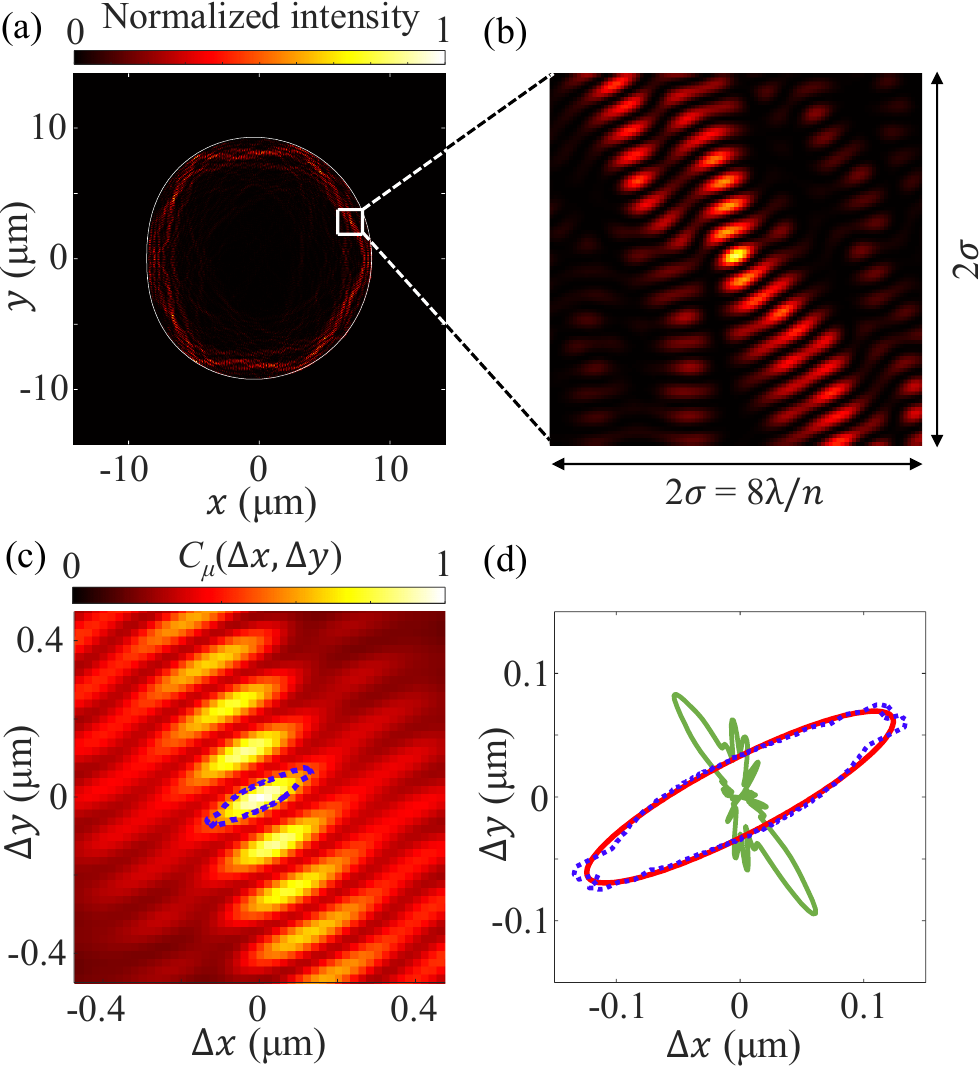}
		\end{center}
		\caption{Local structure size of a cavity resonance. (a)~Calculated intensity distribution of a typical high-$Q$ mode in a \Lim cavity with rough boundary ($\lambda = 798.6$~nm, Q = $1.05\times10^4$). (b)~Magnification of a $2\sigma \times 2\sigma$ region [white box in (a)], where $2\sigma = 8\lambda/n = 1.9~\mu$m is the full width of the wavelet in Fig.~\ref{fig:wavelet}(b). (c)~Spatial intensity correlation function $C_{\mu}(\Delta \vec{r}; \vec{r}_0)$ in the magnified region. The blue dotted line indicates the half-maximum contour line where $C_{\mu} = 1/2$. (d)~The half-maximum contour line (blue dotted line) is fitted by an ellipse (solid red line) with a major (minor) axis of $0.28~\mu$m ($0.06~\mu$m). The solid green line indicates the directionality diagram $|w_\mu(\vec{r}_0, \theta)|$ in a polar plot. Its maximum at $\theta = 123^\circ$ gives the local wave propagation direction, consistent with (b).}
		\label{fig:structSize}
	\end{figure}
	
	The calculation of the local structure size of a lasing mode is illustrated by the example in Fig.~\ref{fig:structSize}. Figure~\ref{fig:structSize}(a) shows the intensity distribution of a high-$Q$ mode of a \Lim resonator with surface roughness, and Fig.~\ref{fig:structSize}(b) is a magnification of it in a local area (white box). It reveals a few standing waves in the direction $\sim 120^o$. The feature size in the direction transverse to propagation, often called the transverse wavelength, is obtained from the spatial intensity correlation function of mode $\mu$ around a point $\vec{r}_0 = (x_0, y_0)$,
	\begin{equation} 
        \label{eq:spmufun} C_{\mu}(\Delta \vec{r}; \vec{r}_0) = \int \limits_{x_0 - \delta x}^{x_0 + \delta x} dx \int \limits_{y_0 - \delta y}^{y_0 + \delta y} dy \, I(\vec{r}) I(\vec{r} + \Delta \vec{r}) \, , 
    \end{equation}
	where $I(\vec{r}) = |\Psi(\vec{r})|^2$ is the intensity distribution of the mode, $\Delta \vec{r} = (\Delta x, \Delta y)$, and the correlation function is normalized to $C_{\mu}(\vec{0}) = 1$. We note that the integration is over a $2\delta x \times 2\delta y = 2\sigma \times 2\sigma$ large region around $\vec{r}_0$, where $2\sigma=8\lambda/n$ = 1.9 $\mu$m is the full width of the wavelet we used. 
	
	The spatial intensity correlation function is shown in Fig.~\ref{fig:structSize}(c). The central maximum of the function can have a circular or an elongated shape, depending on whether the local intensity structure is more isotropic or anisotropic. Its shape, size, and orientation are a good measure of the fine structure of the intensity distribution. We determine the structure size by extracting the half-maximum contour line where $C_\mathrm{\mu} = 1/2$ around $\Delta \vec{r} = (0, 0)$ [blue dotted lines in Fig.~\ref{fig:structSize}(c)]. This contour line has an approximately elliptic shape, and it is plotted together with an ellipse fit (solid red line) in Fig.~\ref{fig:structSize}(d). The fitted ellipse has a major (minor) axis of $0.28~\mu$m ($0.06~\mu$m) oriented in the direction $\theta = 118^\circ$ ($28^\circ$). The transverse wavelength is given by the major diameter of the fitted ellipse, which is perpendicular to the local wave propagation direction as shown below. 
	
	The orientation of the ellipse is dictated by the local directionality, as shown by a polar plot of the directionality diagram $|w_\mu(\vec{r}_0, \theta)|$ (obtained from the wavelet transform) superimposed in Fig.~\ref{fig:structSize}(d) as a solid green line. Its maximum is at $\theta = 123^\circ$, which agrees well with the orientation of the minor ellipse axis. Thus, it represents the local direction of wave propagation, and the major ellipse axis $s_\mathrm{l}$ yields the transverse wavelength. If it is much larger than the in-medium wavelength, optical lensing effects may cause light to focus in the propagation direction, thus forming spatial filaments that can lead to lasing instabilities. 
	
	For every lasing mode $\mu$, we repeat the above procedure for every location $\vec{r}$ inside the resonator and obtain $\tilde{s}_{\mu}(\vec{r})$. The spatially-resolved structure size of a microlaser is obtained by averaging over all lasing modes, weighted by their power $P_{\mu}$ (obtained from the SPA-SALT simulation):
	\begin{equation} 
	s(\vec{r}) = \frac{ \sum_{\mu} P_{\mu} \tilde{s}_{\mu}(\vec{r}) }{ \sum_{\mu} P_{\mu} } \, , 
	\end{equation}
	Finally, we average the local structure size over the entire cavity area, again weighted by local intensity $I(\vec{r})$ [averaged intensity of all lasing modes, Eq.~(\ref{eq:intensity})],
	\begin{equation} 
	\langle s \rangle = \frac{ \int d\vec{r} s(\vec{r}) I(\vec{r}) }{ \int d\vec{r} I(\vec{r}) } \, , 
	\end{equation}
	The mean structure size $\langle s \rangle$ for each of the five cavity shapes is presented in Fig.~3(b) of the main text. 

	\subsection{Ray dynamics}
	
	\subsubsection{Cavity geometry}
	
	In this section, we discuss in detail the ray dynamical properties of the five cavity geometries: D-shape, stadium, \Lim, ellipse, and square.
	
	The D-cavity is a circle of which a section has been cut off, where the distance from the circle center to the cut is $R_D/2$, where $R_D$ is the radius of the cavity. The stadium consists of a square with side length $a$ between two semicircles of radius $a/2$. Both the D-cavity and stadium have completely chaotic ray dynamics \cite{Bunimovich1979, Ree1999} with the exception of marginally unstable periodic orbits with measure zero in phase space. Their classical trajectories cover the whole billiard and are spatially extended, but due to the leakiness of the dielectric cavity, the long-lived trajectories feature central regions of reduced intensity~\cite{Bittner2020}.
	
	The boundary of the \Lim cavity \cite{Robnik1983} is defined in polar coordinates by $r(\varphi) = R_0 (1 + \epsilon \cos\varphi)$, where $\varphi$ is the azimuthal angle, $R_0$ is the mean radius (averaged over $\varphi$), and $\epsilon = 0.42$ is the deformation parameter. In contrast to the D-cavity and stadium, its ray dynamics is predominantly, but not completely chaotic: since it is a convex cavity, there remain invariant surfaces with whispering-gallery type trajectories very close to its boundary \cite{Lazutkin1973, Wiersig2008}, and there is a small stable island with integrable ray dynamics around the horizontal diameter orbit \cite{Robnik1983, Wiersig2008}, as well as further tiny stable islands around other stable periodic orbits \cite{Dullin2001}. So technically it is a mixed system, but we will consider it a chaotic billiard, since the integrable regions of phase space are too small to support any localized resonances for the cavity size considered here. The long-lived trajectories are concentrated at the cavity boundary and confined by total internal reflection. Even though the chaotic trajectories eventually explore the whole phase space, in an open dielectric cavity, they will quickly lose their intensity once they reach the central region of the cavity, as their incident angles onto the boundary are below the critical angle for total internal reflection.

	Our elliptic cavities have the aspect ratio $b/a = 2$, where $b$ and $a$ represent the major and minor axes of the ellipse. Like the circle, the ellipse has completely integrable ray dynamics \cite{Berry1981}. The long-lived trajectories are spatially localized near the boundary and are confined by total internal reflection. While there exists another type of modes based on so-called librator trajectories \cite{Keller1960}, these have a much shorter lifetime and thus cannot lase at the pump levels considered in this study.
	
	Finally, the square has completely integrable ray dynamics like the ellipse, however, the structure of ray trajectories is quite different. Since the moduli of the momentum components, $|k_x|$ and $|k_y|$, are conserved quantities, there exist trajectories confined by total internal reflection at every reflection, which give rise to modes with extremely long lifetimes. The trajectories exhibit a very regular structure like for the ellipse, however, they cover the whole cavity homogeneously in contrast to the highly localized trajectories of the ellipse.

	\subsubsection{Ray tracing algorithm}
	
	We perform ray tracing for all five cavity shapes with smooth boundaries. We set the radius $R_D$ of the D-shaped cavity to unity, and the sizes of the other four cavity shapes are chosen such that they have the same cavity area ($\simeq 2.53\times R_D^2$). We consider the transverse-electric (TE) polarization of light, to be consistent with the polarization of edge emission from GaAs quantum-well lasers. The effective refractive index inside the cavity is $\neff=3.37$, and it is $1$ outside.
	
	We launch the optical rays with unit intensity and let them propagate with the in-medium speed of light, $c/\neff$. Since the dielectric billiards we consider here are open systems, the ray intensity can change at reflections on the cavity boundary.  Depending on the angle of incidence with respect to the boundary normal, $\chi \in (-\pi/2, \pi/2)$, the intensity either decreases due to refractive loss ($\sin|\chi| < 1/\neff$) according to the Fresnel reflection coefficients or remains unchanged thanks to total internal reflection ($\sin|\chi| \geq 1/\neff$), ignoring evanescent tunneling. We track the trajectories in space as well as their intensity decay as a function of the actual time of flight (instead of the number of reflections). 
	
	The aim of the ray tracing simulations is to calculate the local directionality analogously to the case of the wave simulations discussed above in order to demonstrate the correspondence of rays and waves in this context. This requires averaging over specific ensembles of rays, which are different for chaotic and integrable billiards, respectively. The appropriate ensembles can be chosen via the initial conditions of the rays, as explained below.
	
	For cavities with chaotic ray dynamics (D-cavity, stadium, and \Lim), all high-Q resonances correspond to the same ensemble of long-lived trajectories \cite{Altmann2013, ketzmerick2022chaotic}. We launch an ensemble of $10^6$ trajectories at uniformly distributed random positions on the cavity boundary into uniformly distributed random initial directions $\sin\chi \in (-1,1)$. In order to select the most long-lived trajectories, we sample the rays during the time interval in which the averaged ray intensity exhibits exponential decay after initial transients \cite{Bittner2020}. We choose the time intervals of $[10.0, 30.0]$, $[11.9, 35.7]$, and $[43.0, 60.2]$ for D-cavity, stadium, and \Lim, respectively, where the unit of time is $\neff R_D/c$. 
	
	For integrable cavities (ellipse and square), each individual resonance corresponds to one torus in phase space \cite{Berry1977}. We hence launch all rays at a fixed position on the cavity boundary with $10^3$ different angles of incidence $\chi$ such that $\sin\chi$ is evenly spaced in $(-1,1)$. Then each ray propagates on a different torus in phase space and explores its torus entirely given enough time\footnote{Except for periodic trajectories, which we exclude here.}. Here, we only sample the ray trajectories that are permanently confined by total internal reflection since the decaying trajectories correspond to low-$Q$ modes that are irrelevant for lasing. The time interval for sampling rays is $[0, 10^4] \neff R_D/c$, which is sufficiently long to homogeneously sample the tori.
	
	A small modification of the initial conditions is needed for the square. Due to its symmetry with respect to the diagonal axes, a resonance of a square resonator exhibits eight plane-wave components with wave vectors $(k_x=\pm k_1, k_y =\pm k_2)$ and $(k_x=\pm k_2, k_y =\pm k_1)$~\cite{Bittner2013b}. However, a single ray trajectory in a square billiard represents only four of these eight propagation directions. Therefore, we combine two trajectories with initial angles of incidence of $\chi$ and $\chi+\pi/2$ to represent a single resonance of the square cavity, where $\chi \in (-\pi/2, 0)$.

	\begin{figure*}[ht]
		\begin{center}
			\includegraphics[width = 18 cm]{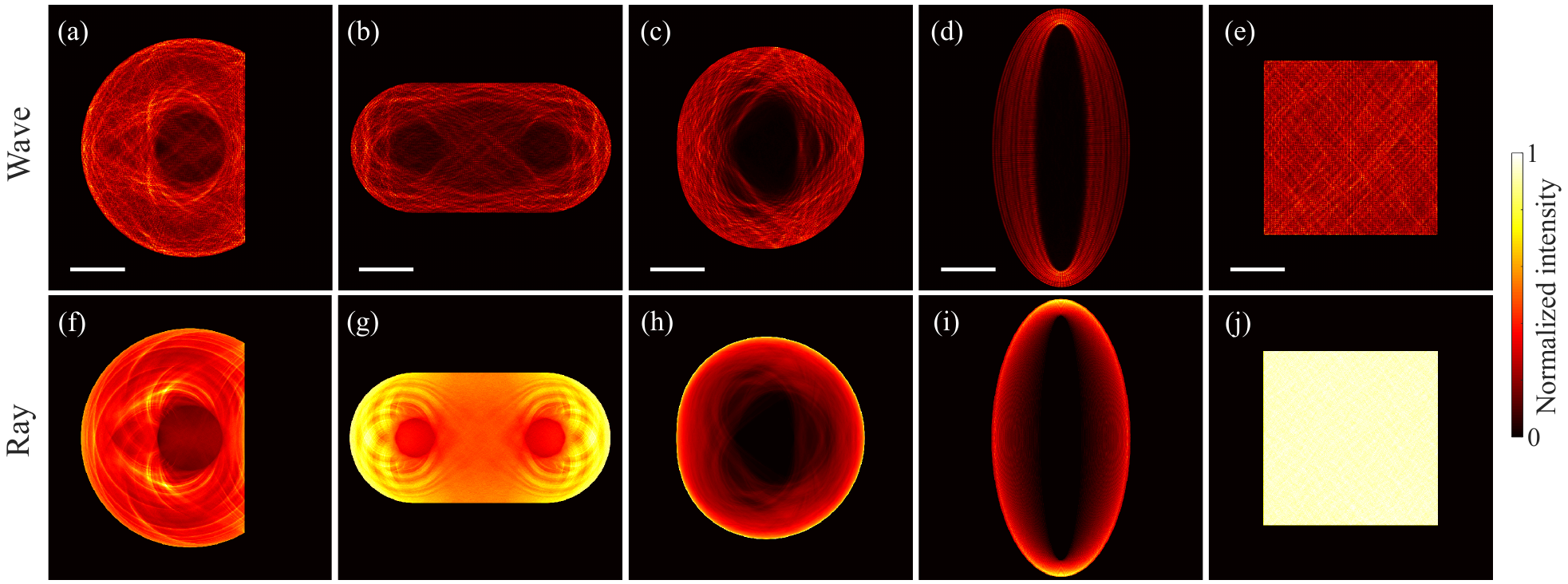}
		\end{center}
		\caption{Spatial intensity distributions of lasing modes. The top row is obtained from wave simulations for (a)~D-cavity, (b)~stadium, (c)~\Lim, (d)~ellipse, and (e)~square cavities with surface roughness. The field intensity is averaged over all lasing modes in three cavities with different surface roughness realizations that are statistically equivalent and described in subsection \ref{sec:rough}. The scale bars are $5~\mu$m long. The bottom row (f-j) represents the ray intensity distribution from ray tracing simulations with smooth cavity boundaries. The spatial structures inside the cavity show excellent agreement between wave and ray simulations.}
		\label{fig:wfavg}
	\end{figure*}

	\subsubsection{Local directionality}
	
	From all sampled ray trajectories, we sum the intensities of all rays with direction $\theta$ inside a local area centered at $\vec{r}$ to obtain $R(\vec{r},\theta)$, where $\vec{r}$ is the position inside the cavity, and $\theta$ is the direction of ray propagation with respect to the horizontal axis. The bin size for sampling the direction $\theta$ is 1$^{\circ}$. The radius of the local area is chosen as $(9.5 \times 10^{-2}) R_D$, so that when $R_D$ is 10 $\mu$m as in the wave simulations, this radius becomes 0.95 $\mu$m, which is exactly the half-width of the wavelet used for wave simulations (see Fig.~\ref{fig:wavelet}). 
	
	The local angular spread of ray propagation directions is defined as
	\begin{equation}
	\sigma_R(\vec{r})=\min_{\theta_0}\sqrt{ \frac{ \int^{+\frac{\pi}{2}}_{-\frac{\pi}{2}}\theta^2[R(\vec{r},\theta+\theta_0) + R(\vec{r},\theta+\theta_0-\pi)]d\theta }{ \int^{+\pi}_{-\pi}R(\vec{r},\theta)d\theta } },
	\end{equation}
	which is analogous to Eq.~(\ref{eq:angWave}) for the wave simulations. The local directionality is defined by the inverse of the angular spread,
	\begin{equation}
	D_R(\vec{r}) = \frac{1}{\sigma_R(\vec{r})}.
	\end{equation}
	The maps of $D_R(\vec{r})$ are presented in Figs.~4(e$_1$-e$_5$) of the main text, showing an excellent agreement with the local directionality maps $D_W(\vec{r})$ obtained from wave simulations.
	
	Finally, we calculate the mean directionality, averaged over the entire cavity area
	\begin{equation}
	\langle D_R \rangle = \frac{\int d\vec{r} D_R(\vec{r}) I(\vec{r})}{\int d\vec{r} I(\vec{r})},
	\end{equation}
	where $I(\vec{r}) = \int R(\vec{r},\theta)d\theta$ is the ray intensity distribution inside the cavity. The calculated $\langle D_R \rangle$ for the five different cavity shapes, as well as the comparison with $\langle D_W \rangle$, are presented in Fig.~3(b) of the main text.

	\subsubsection{Spatial intensity distributions} \label{sec:wfavg}
	
	Finally, we confirm ray-wave correspondence by comparing the results of ray tracing with wave simulations. To this end, we present the averaged intensity distributions inside the cavity in Fig.~\ref{fig:wfavg}. 
	
	For wave simulations of microcavities with rough boundary (top row), the lasing modes at the pumping level of $10$ times the lasing threshold are calculated using SPA-SALT, and the average intensity distributions are given by $I(\vec{r}) = \sum_{\mu} P_\mu |\Psi_\mu(\vec{r})|^2$, where the lasing modes are weighted by their power $P_\mu$. We also averaged the intensity distributions over three surface roughness realizations. 
	
	For ray simulations (bottom row), we show $I(\vec{r}) = \int R(\vec{r},\theta)d\theta$. Here we use a finer spatial resolution of $(1.2 \times 10^{-3})R_D$, in order to reveal detailed features in the intensity profiles inside a cavity.
	
	The ray and wave simulations yield qualitatively consistent intensity profiles. The D-cavity and stadium have a quite homogeneous average intensity distribution, but we observe central regions of lower intensity due to the leakage at the dielectric cavity boundaries \cite{Bittner2020}. The average intensity distributions of the \Lim and ellipse resonators are localized at the boundary due to the prevalence of whispering gallery modes. Thus they are highly localized, though the boundary roughness in the wave simulations increases the area they cover compared to ray simulations. The ray intensity distribution in the square with a smooth boundary is perfectly homogeneous over the whole cavity. However, the rough boundary in wave simulation leads to the formation of diagonal stripes of higher intensity, rendering the wave intensity distribution a bit less homogeneous compared to the ray intensity. In general, the ray tracing simulations in 2D cavities with smooth boundaries correctly predict all the main characteristics of the spatial structure of lasing modes in wave simulations of cavities with small boundary roughness. Considering the huge amount of computational resources required for wave simulations, ray tracing can be an efficient tool to predict the lasing characteristics of realistic microcavities. 

\end{document}